\documentclass[useAMS,usenatbib,referee]{biom}
\def\bSig\mathbf{\Sigma}

\usepackage[english]{babel}
\usepackage[utf8]{inputenc}
\usepackage{graphicx}
\usepackage{todonotes}
\usepackage{enumerate}
\usepackage{amsmath,amsfonts}
\usepackage{subfig}
\usepackage[colorlinks,citecolor=blue,urlcolor=blue]{hyperref}
\usepackage{enumitem}

\title[Adaptive Bayesian Spectral Analysis of Nonstationary Biomedical Time Series]{Adaptive Bayesian Spectral Analysis of Nonstationary Biomedical Time Series}

\author{Scott A. Bruce$^{1}$, 
Martica H. Hall$^{2}$, Daniel J. Buysse$^{2}$, and Robert T. Krafty$^{3,*}$\email{rkrafty@pitt.edu} \\
$^{1}$Department of Statistical Science, Temple University, Philadelphia, Pennsylvania, 19122, U.S.A. \\
$^{2}$Department of Psychiatry, University of Pittsburgh, Pittsburgh, Pennsylvania, 15213, U.S.A. \\
$^{3}$Department of Biostatistics, University of Pittsburgh, Pittsburgh, Pennsylvania, 15261, U.S.A.}

\begin{document}


\date{{\it Received September} 2016. {\it Revised January} 2017.  {\it
Accepted February} 2017.}



\pagerange{\pageref{firstpage}--\pageref{lastpage}} 
\volume{64}
\pubyear{2017}
\artmonth{February}


\doi{10.1111/j.1541-0420.2005.00454.x}


\label{firstpage}


\begin{abstract}
Many studies of biomedical time series signals aim to measure the association between frequency-domain properties of time series and clinical and behavioral covariates.  However, the time-varying dynamics of these associations are largely ignored due to a lack of methods that can assess the changing nature of the relationship through time.  This article introduces a method for the simultaneous and automatic analysis of the association between the time-varying power spectrum and covariates. The procedure adaptively partitions the grid of time and covariate values into an unknown number of approximately stationary blocks and nonparametrically estimates local spectra within blocks through penalized splines.  The approach is formulated in a fully Bayesian framework, in which the number and locations of partition points are random, and fit using reversible jump Markov chain Monte Carlo techniques. Estimation and inference averaged over the distribution of partitions allows for the accurate analysis of spectra with both smooth and abrupt changes. The proposed methodology is used to analyze the association between the time-varying spectrum of heart rate variability and self-reported sleep quality in a study of older adults serving as the primary caregiver for their ill spouse.

\end{abstract}

%



\maketitle


%

\section{Introduction}
\label{sec:intro}

The frequency-domain properties of many biological time series signals have been found to contain valuable information.  Heart rate variability \citep{HRVstandards} and electroencephalography \citep{EEGbook} are examples of signals where studies in the frequency domain have uncovered interpretable physiological information.  As a result, many biomedical studies collect and analyze time series signals from multiple subjects to better understand how power spectra relate to clinical and behavioral variables.  The dynamic nature of most biological processes means that these time series are rarely stationary.  A method that can quantify the association between time-varying spectra and study covariates is needed to properly reflect the temporally-evolving nature of the relationship and to better understand dynamic biological processes.

The motivating study for this article seeks to quantify the association between psychological stress and self-reported sleep quality in older adults who are the primary caregiver for their spouse.  In this study, heart rate variability (HRV) is observed during a night of sleep.   HRV is a measure of the variability in time elapsed between consecutive heart beats.  The power spectrum of HRV provides noninvasive, indirect measures of autonomic nervous system activity \citep{HRVstandards} and has been used in this capacity to obtain objective measures of stress and arousal \citep{hall2004}.  Participants also completed a self-reported questionnaire used to compute the Pittsburgh Sleep Quality Index (PSQI) \citep{psqi}, a clinically validated summary measure of sleep quality.  In quantifying the association between the time-varying spectrum of HRV and PSQI, we aim to better understand the dynamic relationship between stress levels and sleep quality.

In the clinical and biological literature, such data are commonly analyzed using an ad hoc, four stage procedure.  In the first stage, time series are partitioned into prespecified, equally-spaced intervals (e.g. two minute epochs). In the second stage, collapsed measures within prespecified frequency bands that quantify specific characteristics of heart rate variability are computed for each interval \citep{BurrHRV}.  For example, normalized high frequency (HFnu) is calculated as the total power from high frequencies (0.15-0.4 Hz) divided by total power from high and low frequencies (0.04-0.4 Hz) and has been found to be inversely related to experimental stress \citep{hall2004}. In the third stage,  interval specific measures are averaged across time to obtain  time-invariant measures for each subject.  Finally, in the last stage, the association between these temporally averaged measures and clinical and behavioral variables are assessed using standard tools such as Pearson's correlations and ANOVA \citep{HRVstandards, hall2004}.  There are several serious drawbacks to this ad hoc approach. First, time series are segmented a priori into intervals without regard for the dynamics of the series.  Second, spectral measures of a given subject are averaged over intervals to obtain time-invariant measures. This inhibits the ability to assess dynamic relationships between spectral measures and other study variables.  Lastly, the final stage treats the estimated spectral measures as known parameters, thus ignoring the variability accumulated through the multiple stages of the estimation procedure and leading to inaccurate inference.

In the statistics literature, many models and methods for the spectral analysis of nonstationary time series have been studied.   These include methods with piecewise stationary estimators \citep{adak,slex2005,davis2006}, methods that smooth across time to obtain slowly-varying estimators \citep{daulhaus,guo2003,qinwang2009}, and Bayesian methods that provide estimators that can adapt to both abrupt and slowly-varying temporal dynamics \citep{adaptspec,yang2016}.  However, all of these methods follow traditional time series analysis focusing on analyzing a single time series. While there are many methods for the time-frequency analysis of a single time series, methods for the analysis of a collection of time series from multiple subjects whose spectra are associated with study covariates are few and preliminary. Proposals by \citet{qinguolitt2015} and \citet{fiecas2016} enable the analysis between time-dependent spectra and covariates when spectra evolve continuously and smoothly as functions of both time and  covariate.  In many applications, the spectral properties of a time series may change suddenly.  For instance, in our motivating study, there might be smooth trends in HRV spectra throughout the night, but abrupt changes are expected when the body transitions between different sleep stages.  Furthermore, there is evidence that the relationship between PSQI and physiology may show abrupt threshold effects \citep{krafty2012}.
There currently are no formal statistical methods that allow for the analysis of associations between covariate and time-varying power spectra that can capture both abrupt and slowly varying dynamics.

This article proposes a method for simultaneous and automatic estimation of the association between the time-varying spectrum and covariates.  A
covariate-indexed locally-stationary model is presented, in which spectra are functions of frequency, covariate, and time; local spectra are nonparametrically estimated using penalized splines.  The model is formulated in a Bayesian framework where the number and location of time and covariate indices are random variables and fit using reversible-jump Markov chain Monte Carlo (RJMCMC) techniques.  The proposal provides a flexible and adaptive estimator of the time- and covariate-varying spectrum that broadens the existing scope of processes and scientific questions that can be addressed in three important ways.  First, the approach uses the data to determine the appropriate number and location of time and covariate indices rather than requiring a prespecified segmentation scheme for the locally-stationary model. Second, by averaging over the posterior distribution of the number and locations of time and covariate indices, the estimator can recover both smooth and abrupt changes in the spectrum across time and covariate.  Finally, the sampling procedure naturally enables inference on any function of the time and covariate-varying spectrum, including collapsed measures through time and the location and magnitude of abrupt changes, both of which are of importance in applied analyses.

This paper is organized as follows.  Section \ref{sec:motivatingstudy} provides more details about the motivating study: the AgeWise Caregiver Study.  Section \ref{sec:method} provides a brief methodological background on
developments in frequency-domain time series analysis.  Section \ref{sec:proposedmethod} introduces the proposed Bayesian covariate-indexed locally-stationary model along with the sampling and estimation procedure.  Illustrations using simulated slowly and abruptly changing spectra are presented in Section \ref{sec:simulation}, and the proposed method is applied to the AgeWise Caregiver Study in Section \ref{sec:application}.  Concluding remarks are found in Section \ref{sec:discussion}.

\section{Motivating Study}
\label{sec:motivatingstudy}

Older adults who are the primary caregiver for their ill spouse often experience significant mental and emotional stress, and are likely to develop some form of sleep disturbance that can negatively impact health and functioning \citep{sleepdementia}.  A goal of the AgeWise Caregiver Study conducted at the University of Pittsburgh was to gain a better understanding of the association between stress and sleep in older adults who were the primary caregiver for their ill spouse to inform the development of behavioral interventions to enhance their sleep.

The current analysis considers data from 30 men and women 60-89 years of age.  Each participant served as the primary caregiver for their spouse who was suffering from a progressive dementing illness such as Alzheimer’s or advanced Parkinson’s disease.   As previously discussed, study participants completed a self-reported questionnaire used  to formulate a PSQI score \citep{psqi}.  The PSQI score is a popular clinical measure of self-reported sleep disturbances and of how these disturbances affect daily functioning over a one-month period.  PSQI scores can range in value from 0-21; larger scores represent more disturbed sleep, with scores of 6 or larger typically taken as an indicator of clinically disturbed sleep.  In our sample, PSQI scores ranged from 1-13 and had a mean of 7. 

Participants were also studied during a night of in-home sleep through ambulatory polysomnography (PSG), which is the comprehensive recording of electrophysiological changes during sleep. The PSG used in the study included an electrocardiograph (ECG) to monitor heart activity.  The ECG was used to locate the timing of heart beats, which were then differenced, detrended, linearly interpolated, and resampled at 1 Hz to compute a HRV series continuously throughout the night. During the night, the body cycles through two types of sleep:  non-rapid eye movement (NREM) sleep, which contains deep sleep, and rapid eye movement (REM) sleep, in which dreaming typically occurs. We isolated a 10 minute long epoch of HRV for each participant during the 5 minutes before and 5 minutes after the first onset of REM sleep.

\begin{figure}[h]%
\centering
\subfloat{\includegraphics[height=2in]{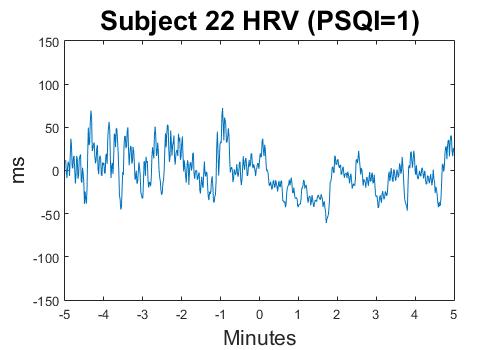}}%
\qquad
\subfloat{\includegraphics[height=2in]{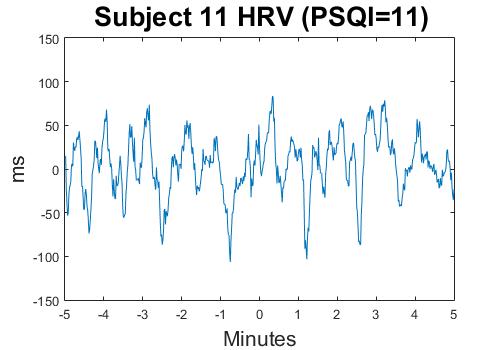}}
\caption{Demeaned heart rate variability time series for 5 minutes before and 5 minutes after first onset of REM sleep for two study participants.  PSQI scores for the participants are 1 and 11, respectively, indicating worsening sleep quality.}
\label{fig:data}
\end{figure}

In healthy individuals, the spectrum of HRV is expected to change as subjects move from NREM to REM  \citep{healthyHRV}.  Disruptions in these patterns have been observed under stress and represent a mechanism through which stress may negatively affect health and functioning \citep{hall2004}.  We desire an analysis of the association between the time-varying spectrum of HRV and PSQI to quantify how the dynamics of HRV spectra during the transition from NREM to REM are associated with sleep quality. HRV time series and PSQI scores for two subjects are displayed in Figure \ref{fig:data} to illustrate study data.  HRV time series and PSQI scores for all participants can be found in online supplementary materials (see \hyperref[sec:websuppA]{Web Supplement A}).

\section{Methodological Background, Single Time Series}
\label{sec:method}
To motivate the proposed approach to analyzing associations between covariates and time-varying spectra from replicated time series, we first provide a brief explanation of the technique for estimating the power spectrum of a single time series.

\subsection{Stationary Time Series}
\label{sec:stationary}

For a zero-mean stationary time series, $\{X_t\}$, $t=0, \pm 1, \pm 2, \ldots, $ the frequency domain characteristics can be obtained from its spectral representation \citep{cramer42},
\begin{equation*}
X_t = \int_{-1/2}^{1/2}A(\nu)\exp(2\pi i t \nu) dZ(\nu),
\label{eqn:specrep}
\end{equation*}
where $A(\nu)$, known as the transfer function, is a complex-valued function that is periodic $A(\nu) = A(\nu + 2 \pi)$ and Hermitian $A(\nu) = \overline{A(-\nu)}$, and where $Z(\nu)$ is a zero-mean orthogonal process such that
\begin{equation*}
\mathrm{cov}\left[dZ(\nu), dZ(\nu')\right] = \begin{cases}
									0 \quad \mathrm{if} \; \nu \neq \nu' \\
                                    d\nu \quad \mathrm{otherwise.}
                                  \end{cases}
\end{equation*}

The power spectrum
of the series is defined as
\begin{equation*}
f(\nu) = |A(\nu)|^2 = \sum_{k=-\infty}^{\infty} \gamma_k\exp(-2\pi i \nu k)
, \; \nu \in \mathbb{R},
\label{eqn:specdef}
\end{equation*}
where $\gamma_k= \mathrm{cov}(X_t,X_{t+k}) $ is the autocovariance function of $\{X_t\}$.  We can also recover $\gamma_k$ by applying the inverse Fourier transform




\begin{equation*}
\gamma_k = \int_{-1/2}^{1/2} f(\nu) \exp(2\pi i \nu k) d\nu.
\label{eqn:ifft}
\end{equation*}

\noindent Note that the variance of the time series, $\gamma_0$, can be found to be
\begin{equation*}
\gamma_0 = \int_{-1/2}^{1/2} f(\nu) d\nu.
\end{equation*}
In other words, the power spectrum can be interpreted as the decomposition of the variance over frequencies \citep{wei} and is the primary tool used to describe the frequency-domain properties of stationary time series.

Suppose that $\{X_t\}$ has a bounded positive power spectrum. Given a realization, $x_1,\ldots,x_T$, the periodogram at frequency $\nu$ is
\begin{equation*}
Y(\nu) = \frac{1}{T}\left| \sum_{t=1}^T x_t\exp(-2\pi i \nu t) \right|^2.
\label{eqn:periodogram}
\end{equation*}
While the periodogram is an unbiased estimator of the true spectrum, the asymptotic variance does not tend to zero as sample size, $T$, increases \citep{wei}, so smoothed estimators that share information across frequencies are considered.  Efficient estimators can be obtained by utilizing the large sample distribution of the periodogram \citep{pawitan1994}.  Let $n=\lfloor T/2 \rfloor - 1$ and $\nu_k=k/T$ for $k=1,\ldots,n$ be the Fourier frequencies.  \citet{whittle} showed that under appropriate conditions, for large $T$, the likelihood of $\mathbf{x}=(x_1,\ldots,x_{T})'$ given $\mathbf{f}=[f(\nu_1),\ldots,f(\nu_n)]'$ can be approximated by
\begin{equation*}
p(\mathbf{x}|\mathbf{f}) \approx (2\pi)^{- n/2 }\prod_{k=1}^{n} \exp\left\{-
[\log f(\nu_k) + Y(\nu_k)/f(\nu_k)]\right\}.
\label{eqn:whittle}
\end{equation*}
%

Since $f$ is a positive function, to enable unconstrained estimation, the spectrum is modeled on the logarithmic scale \citep{wahba1980}.  
We consider a Bayesian penalized spline model for $f$; a thorough discussion of penalized spline models and their Bayesian formulation can be found in \citet{ruppert2003}.  Linear splines are used as they have been found to be more accurate in capturing changes in the power spectra compared to higher order splines \citep{adaptspec}.  The log spectrum is modeled as
\[ \log f(\nu) \approx \alpha + \sum_{b=1}^{B} \beta_b \cos(2 \pi b \nu).\]
 The functions $\cos(2 \pi b \nu)$ are the Demmler-Reinsch basis for linear, periodic, even splines when data are observed on an evenly spaced grid (i.e. the Fourier frequencies) \citep{schwarz2016}.  We define the $n \times B$ matrix of the basis functions evaluated at the Fourier frequencies as 
 $\mathbf{Z}$ where $\{Z\}_{i,b} = \cos(2 \pi b \nu_i)$.  
   Prior distributions are assumed on the coefficients such that $\boldsymbol \beta = \left( \beta_1, \dots, \beta_B \right)' \sim N(0,\tau^2D_B)$, where $D_B=\mathrm{diag}(\{2\pi b\}^{-2})$, which is independent of $\alpha \sim N(0, \sigma^2_{\alpha})$. The smoothing parameter $\tau$ controls the roughness of the log spectrum such that, as $\tau \rightarrow 0$, the log spectrum tends to a constant function of frequency with probability 1.   A uniform prior is placed on $\tau^2$ such that $p(\tau^2)= 1/\tau^2$.  The hyperparameter $\sigma_{\alpha}^2$ is set to be a large, fixed number. 

A two-step sampling scheme for the parameters $\alpha, \boldsymbol{\beta},$ and $\tau^2$ using MCMC methods can be constructed as follows \citep{adaptspec}.
\begin{enumerate}
\item Given a realization of the log periodogram, $\log(\mathbf{Y}) = \{\log[Y(\nu_1)],\ldots,$ $\log[Y(\nu_{n})]\}',$ and basis functions $\mathbf{Z}$, $\alpha$ and $\boldsymbol{\beta}$ are sampled jointly in a Metropolis-Hastings (M-H) step from
\begin{equation}\label{eqn:sampstat1}
\begin{split}
p(\alpha,\boldsymbol{\beta}|\tau^2,\log(\mathbf{Y}),\mathbf{Z})
\propto \exp &\left\{-
\sum_{k=1}^{n}[\alpha + \mathbf{z}_k'\boldsymbol{\beta} + \exp(\log[Y(\nu_k)] - \alpha - \mathbf{z}_k'\boldsymbol{\beta})] \right. \\
&\left. - \frac{\alpha}{2\sigma_a^2} - \frac{1}{2\tau^2}\boldsymbol{\beta}'D_B^{-1}\boldsymbol{\beta}\right\},
\end{split}
\end{equation}
where $\mathbf{z}_k'$ is the $k$th row of $\mathbf{Z}$.

\item $\tau^2$ is sampled from the inverse gamma distribution with density

\begin{equation}\label{eqn:sampstat2}
p(\tau^2|\boldsymbol{\beta}) \propto (\tau^2)^{-B/2}\exp\left(-\frac{1}{2\tau^2}\boldsymbol{\beta}'D_B^{-1}\boldsymbol{\beta} \right).
\end{equation}
\end{enumerate}

\subsection{Nonstationary Time Series}
 \label{sec:nonstationary}
A model for the spectrum analysis of nonstationary time series can be defined by allowing the transfer function in the Cram\'{e}r spectral representation to vary with time \citep{priestley1965}.
A time-varying transfer function $A(u,\nu)$ is a function of scaled time $u\in [0,1]$ and frequency $\nu \in \mathbb{R}$ such that for every scaled time point $u$, $A(u,\nu) = A(u, \nu + 2 \pi)$ and $A(u, \nu) = \overline{A(u, -\nu)}$.  We consider the class of locally stationary time series of length $T$ such that $X_{t}(t=1,\ldots,T)$ can be represented as
\[X_{t} = \int_{-1/2}^{1/2} A(t/T,\nu)\exp(2\pi i t \nu)dZ(\nu)\]
for a time-varying transfer function $A$ and an orthogonal process $Z$.
The primary measure of interest is the time-varying spectrum
\[f(u,\nu) = |A(u,\nu)|^2.\]
This natural extension of the stationary power spectrum provides information about the variability due to oscillations at frequency $\nu$ around time $uT$.

It should be noted that this class of locally stationary time series differs slightly from the class originally proposed by \cite{daulhaus}.  In order to encompass common parametric models for finite $T$,
\cite{daulhaus} defined locally stationary time series through a series of time-varying transfer functions that converge to $A(u, \nu)$.  Since we are concerned with nonparametric estimation, we define locally stationary time series directly through $A(u, \nu)$ in a manner similar to \cite{guo2003}.  Additionally, both \cite{daulhaus} and \cite{guo2003} require $A(u,\nu)$ to be continuous in both $u$ and $\nu$ so that temporal smoothing can be used obtain consistent estimators.  Here, we consider the more flexible model of \cite{adak}, where $A(u, \nu)$ is continuous as a function of $\nu$, but can be discontinuous as a function of $u$.  This flexibility allows for modeling abrupt temporal changes that occur in many real applications, including our motivating application, where the spectrum of HRV can change abruptly when transitioning between different sleep stages.

A locally stationary time series can be can be approximated by a piecewise stationary process
\begin{equation*}
X_{t} \approx \sum_{j=1}^{m}X_t^{(j)}\delta_{j,m}(t) 
\label{eqn:piecewisestat}
\end{equation*}
where $\delta_{j,m}(t) = 1$ if $t/T \in (\xi_{j-1,m}, \xi_{j,m}]$ and is zero otherwise, $\boldsymbol{\xi}_m = (\xi_{0,m},\ldots,\xi_{m,m})'$ form a partition of [0,1] into approximately stationary segments, and $X_t^{(j)}$ are stationary processes \citep{adak}.
Let $T_{j}$ be the number of observations in the $j$th segment.  Also let $n_j=\lfloor T_j/2 \rfloor - 1$ and $\nu_{kj} = k/T_j$ for $k=1,\ldots,n_j$ be the Fourier frequencies for the $j$th segment, and let $Y_{j}(\nu)$ be the local periodogram within the $j$th segment.  The likelihood can then be approximated by a product of local Whittle likelihoods
\begin{equation*}
L(f_{1,m},\ldots,f_{m,m}|\mathbf{x},\boldsymbol{\xi}_m) \approx \\
\prod_{j=1}^m(2\pi)^{-n_j/2}\prod_{k=1}^{n_j}\exp\{-
[\log f_{j,m}(\nu_{kj}) + Y_{j}(\nu_{kj})/f_{j,m}(v_{kj})]\}.
\label{eqn:nonstatlikelihood}
\end{equation*}
Following \cite{adaptspec}, the stationary Bayesian model from Section \ref{sec:stationary} can be extended to this nonstationary model by placing priors on the number of segments, $m$, and the partition, $\boldsymbol{\xi}_m$, in addition to using the priors for the stationary log spectrum for each segment.  This allows the number and location of partition points to be adaptively estimated from the data, so both smooth and abrupt changes to the spectrum can be captured by averaging over the posterior distribution of the segmentation.  A discrete uniform prior is used for the number of segments such that $\mathrm{Pr}(m=k)=1/M \; \mathrm{for} \; k=1,\ldots,M$ for some maximum $M$.  Given the number of segments, a constrained uniform prior is placed on each segment end point to ensure at least $t_{\mathrm{min}}$ observations are contained within each of the segments:
\begin{equation*}\mathrm{Pr}(\boldsymbol{\xi}_m|m) = \prod_{j=1}^{m-1} \mathrm{Pr}(\xi_{j,m}|\xi_{j-1,m},m)\end{equation*}
\noindent where $\mathrm{Pr}(\xi_{j,m}=t|m) = 1/p_{j,m}$ and $p_{j,m}$ is the number of possible locations for the endpoint of the $j$th segment subject to the constraint that at least $t_{\mathrm{min}}$ observations are contained in each of the segments.  This constraint ensures a good approximation to the likelihood within each approximately stationary segment. 

\section{Proposed Methodology, Replicated Time Series}
\label{sec:proposedmethod}

 In the previous section, relevant concepts are provided relating to spectral analysis of a single time series.  However, a large number of biomedical experiments collect time series in conjunction with covariates from multiple subjects.  We use $L$ to denote the number of subjects and let $X_{\ell t}$ and $w_{\ell}$ represent the time series and covariate from the $\ell$th
subject, respectively.  In this setting, researchers are interested in understanding how covariates may modulate spectral characteristics of the biomedical signal under study.  Therefore, a new modeling framework is required that allows spectral characteristics to vary across both time and covariates.  We now propose a method to analyze associations between covariates and time-varying spectra from replicated time series.  First, we provide a definition for the covariate-indexed time-dependent spectrum.  Then, a flexible and adaptive estimation procedure is constructed that captures both smooth and abrupt changes in the spectrum across both time and covariate dimensions.

\subsection{Covariate-Indexed Time-Dependent Spectrum}
To extend the locally stationary model for a single time series to the covariate dependent setting, we utilize a Cram\'{e}r representation in which the transfer function depends not only on time and frequency, but also on covariate. Without loss of generality,  we assume covariate values have been scaled so that $w \in [0,1]$.  A collection of covariate-indexed locally stationary time series $X_{\ell t}(t=1,\ldots,T, \ell=1,\ldots,L)$ with covariates $w_{\ell}$ are defined as
\[X_{\ell t} = \int_{-1/2}^{1/2} A(t/T,w_{\ell},\nu)\exp(2\pi i t \nu)dZ_{\ell}(\nu)\]
where $Z_{\ell}$ are independent and identically distributed orthogonal processes and $A(u, w, \nu)$ is a complex-valued function of scaled time $u\in [0,1]$, covariate $w \in [0,1]$, and frequency $\nu \in \mathbb{R}$ such that $A(u, w, \nu) = A(u, w, \nu + 2 \pi)$ and $A(u, w, \nu) = \overline{A(u, w, -\nu)}$.

The covariate-indexed time-dependent spectrum is then defined as
\[ f(u,w,\nu) = |A(u,w,\nu)|^2. \]
The focus of most applications is on how $f$ depends on the covariate $w$.  For example, in our motivating study, the primary question of interest is on how time-varying power spectra depend on sleep quality.  We assume that $A$, and subsequently the spectrum $f$, are continuous functions of frequency $\nu$, but that they can have a finite number of discontinuities as functions of scaled time $u$ and of the covariate $w$.  This flexibility allows for modeling abrupt changes in time (such as when transitioning between different sleep states) and covariate (such as changes experienced when a covariate passes a clinically significant threshold).

\subsection{Covariate-Indexed Piecewise Stationary Model}
\begin{figure}[t!]
\centering
\includegraphics[scale=.6]{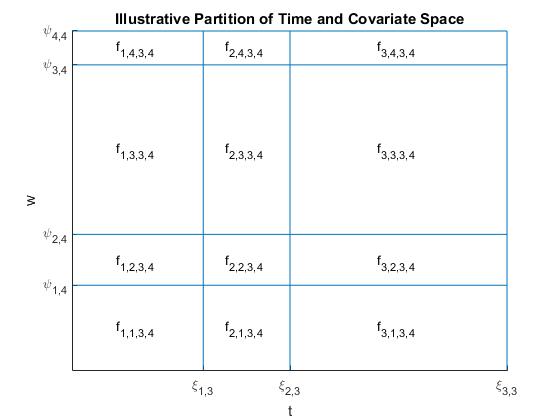}
\caption{Illustration of a partition of the time (t) and covariate (w) space into 3 time segments, $(\xi_{1,3},\xi_{2,3},\xi_{3,3})$ and 4 covariate segments $(\psi_{1,4},\psi_{2,4},\psi_{3,4}, \psi_{4,4})$, and the power spectra for all 12 blocks making up the partition.}
\label{fig:partitionillustration}

\end{figure}

Similar to the locally stationary model discussed in Section \ref{sec:nonstationary}, the covariate-indexed locally stationary model can be approximated by piecewise stationary processes.  Whereas the time dimension was partitioned into approximately stationary segments for the locally stationary model, the plane of time-covariate values is now partitioned into stationary blocks.   After re-scaling time and covariate values to be between 0 and 1, a partition of the time and covariate space into $m$ time-based segments and $p$ covariate-based segments is denoted by $\boldsymbol{\xi}_{m} = (\xi_{0,m},\ldots,\xi_{m,m})'$ and $\boldsymbol{\psi}_{p} = (\psi_{0,p},\ldots,\psi_{p,p})'$ where $\xi_{j,m}$ is the unknown location of the end of the $j$th time segment and $\psi_{g,p}$ is the unknown location of the end of the $g$th covariate segment.  Conditional on $m$, $p$, $\boldsymbol{\xi_m}$, and $\boldsymbol{\psi_p}$,
\begin{equation*}
X_{\ell t} \approx \sum_{j=1}^m X_{\ell t}^{(j,g)}\delta_{j,g,m,p}(t,w_{\ell})
\label{eqn:covpiecewisestat}
\end{equation*}
where for $j=1,\ldots,m$ and $g=1,\ldots,p$, the processes $X_{\ell t}^{(j,g)}$ are stationary with power spectrum $f_{j,g,m,p}(\nu)$, and $\delta_{j,g,m,p}(t,w_{\ell}) = 1$ if $t \in (\xi_{j-1,m}, \xi_{j,m}]$ and $w_{\ell} \in (\psi_{g-1,p}, \psi_{g,p}]$ and is zero otherwise.  See Figure \ref{fig:partitionillustration} for an illustration of a partition of the time and covariate space.

Given a partition of the time and covariate space, ($\boldsymbol{\xi_m}$, $\boldsymbol{\psi_p}$), and letting $Y_{j,\ell}(\nu)$ be the local periodogram for the $\ell$th subject within time block $j$, the likelihood can be approximated by a product of Whittle likelihoods
\begin{align*}
L(f_{1,1,m,p},\ldots,f_{m,p,m,p}|\mathbf{x},\boldsymbol{\xi}_m,\boldsymbol{\psi}_p) \approx \prod_{\ell=1}^L\prod_{j=1}^m(2\pi)^{-n_j/2}\prod_{k=1}^{n_j} \\
\times \; \exp\{-
[\log f_{j,g_{\ell},m,p}(\nu_{kj}) + Y_{j,\ell}(\nu_{kj})/f_{j,g_{\ell},m,p}(\nu_{kj})]\}.
\end{align*}
\subsection{Priors}
For a given number of time segments $m$ and covariate segments $p$ the following set of priors are used.

\begin{enumerate}
\item Priors on the log spectra 
$\log f_{j,g,m,p}(\nu), j=1,\ldots,m, g=1,\ldots,p,$ are assumed to be independent and are as specified in Section \ref{sec:stationary}.

\item Prior on the time partition $\boldsymbol{\xi_m}$ is as specified in Section \ref{sec:nonstationary}.  A similar prior is placed on the covariate partition $\boldsymbol{\psi_p}$:
\begin{equation}\mathrm{Pr}(\boldsymbol{\psi}_p|p) = \prod_{g=1}^{p-1} \mathrm{Pr}(\psi_{g,p}|\psi_{g-1,p},p)\end{equation}
\noindent where $\mathrm{Pr}(\psi_{g,p}=w|p) = 1/p_{g,p}$ and $p_{g,p}$ is the number of possible locations for the endpoint of the $g$th segment and equals the number of unique covariate values greater than the endpoint for the $(g-1)$th segment, $\psi_{g-1,p}$, less the number of remaining segments, $p-g$.
\end{enumerate}

The prior for the number of time segments $m$ is as specified in Section \ref{sec:nonstationary}, and a similar discrete uniform prior is placed on the number of covariate segments $p$: $\mathrm{Pr}(p=g)=1/M_p \; \mathrm{for} \; g=1,\ldots,M_p$ for some maximum $M_p$ to complete the prior specification.

\subsection{Sampling Scheme}
\label{sec:samplingscheme}
Each MCMC iteration conducts two types of moves, within-model moves and between-model moves, that may alter the time and covariate partitions separately and in turn.  A description of the sampling scheme is given here while technical details are provided
in the online supplementary materials (see \hyperref[sec:websuppB]{Web Supplement B}).  We adopt the notation where superscripts $c$ and $p$ denote current and proposed parameter values, respectively.

\subsubsection{Within-Model Moves}
Given the current partition of the time and covariate space, ($\boldsymbol{\xi_{m^c}}$, $\boldsymbol{\psi_{p^c}}$), a single time partition point, $\xi_{k^*,m^c}$, is proposed to be relocated.  Across all covariate segments, the corresponding basis function coefficients in the pair of adjacent time segments impacted by the relocation of the time partition point, $(\boldsymbol{\beta}_{k*,i},\boldsymbol{\beta}_{k*+1,i})$ for $i=1,\ldots,p^c$,
are updated.  These two steps are jointly accepted or rejected in an M-H step.  The smoothing parameters are then updated in a Gibbs step according to Equation (\ref{eqn:sampstat2}).  Then, the same process is carried out to consider relocation of one of the covariate partition points, $\psi_{j^*,p^c}$.  The impacted basis functions, $(\boldsymbol{\beta}_{i,j*},\boldsymbol{\beta}_{i,j*+1})$ for $i=1,\ldots,m^c$, are updated and accepted or rejected in a similar M-H step followed by a similar update to the smoothing parameters.  

\subsubsection{Between-Model Moves}

The number of time segments is either proposed to increase by 1 ($m^p=m^c+1$) or decrease by 1 ($m^p=m^c-1$).

\begin{itemize}
\item If a birth is proposed ($m^p=m^c+1$), a time segment is selected for splitting and an additional partition point is selected from within this segment.  For each of the covariate segments, two new smoothing parameters are created from the current single parameter.  Then, new sets of basis function coefficients are drawn conditional on the new smoothing parameters and the move is accepted or rejected in an M-H step.

\item If a death is proposed ($m^p=m^c-1$), a time partition point is selected for removal.  For each of the covariate segments, a single smoothing parameter is formed from the two currently adjacent smoothing parameters.  Similarly, new sets of basis function coefficients are drawn conditional on the new smoothing parameters and the move is accepted or rejected in an M-H step.
\end{itemize}

This process is then repeated similarly for the covariate partition resulting in a potential birth ($p^p = p^c+1$) or death ($p^p = p^c-1$) in the number of covariate segments. 

 \vspace{-0.5em}

\section{Simulated Examples}
\label{sec:simulation}

To better illustrate the proposed methodology and the flexibility of our approach to capture both smooth and abrupt changes, we present results from a simulated piecewise AR process and a simulated slowly-varying AR process modulated by a covariate.  The model is fit to each simulated data using 10,000 total iterations with the first 2,000 used as a burn-in. The maximum number of time and covariate segments is set to $M_m = M_p =10$, $t_{min}$ is set to 40, and the number of spline basis functions is set to $B=7$.

\subsection{Piecewise AR Process}
We consider a collection of $L=8$ conditional piecewise AR processes of length $T=1000$ with covariates $w_{\ell} = (\ell-1)/(L-1)$ for $\ell=1,\ldots,L$ where
\begin{equation*}
x_{\ell t} =
\begin{cases}
-\phi_{\ell} x_{\ell t-1} + \epsilon_{\ell t}, \qquad \mathrm{for} \; 1 \le t \le 500, \\
+\phi_{\ell} x_{\ell t-1} + \epsilon_{\ell t}, \qquad \mathrm{for} \; 501 \le t \le 1000,
\end{cases}
\label{eqn:abruptdraw}
\end{equation*}

 \begin{equation*}
 \phi_{\ell} =
 \begin{cases}
 0.5 \qquad \mathrm{for} \; 0 \le w_{\ell} \le 0.5 \\
 0.9 \qquad  \mathrm{for} \;  0.5 < w_{\ell} \le 1, \\
 \end{cases}
 \label{eqn:abruptcov}
 \end{equation*}
 
and  $\epsilon_{\ell t} \stackrel{\mathrm{iid}}{\sim} N(0,1)$.
The spectrum has a single abrupt change in time, a single abrupt change in the covariate, and is stationary otherwise.


The posterior distribution of the number and locations of segments can be used to evaluate the presence and location of abrupt changes.  The posterior probability for $m=2$ time segments is estimated as 99.66\% (see Figure \ref{fig:simposttime}), and the posterior probability for $p=2$ covariate segments is 99.96\%.
The posterior mean for the time and covariate partition points are $\hat{\xi}_{1,2}=501$ and $\hat{\psi}_{1,2}=0.4286$.  The covariate value of 0.4286 represents the largest realized value less than the true partition point of 0.5.

\begin{figure}[t!]%
\centering
\subfloat[]{\includegraphics[height=2in]{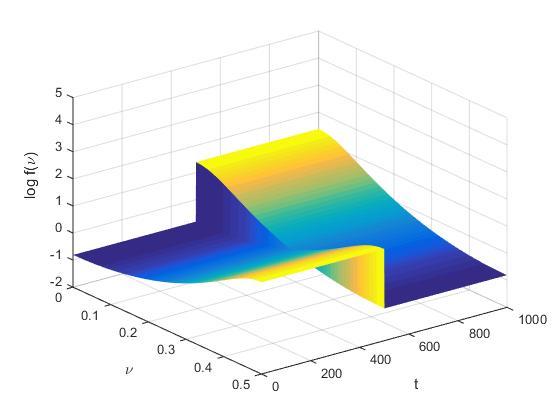}}%
\qquad
\subfloat[]{\includegraphics[height=2in]{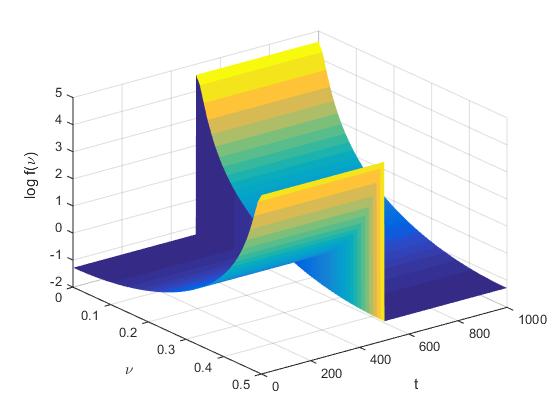}} \\
\subfloat[]{\includegraphics[height=2in]{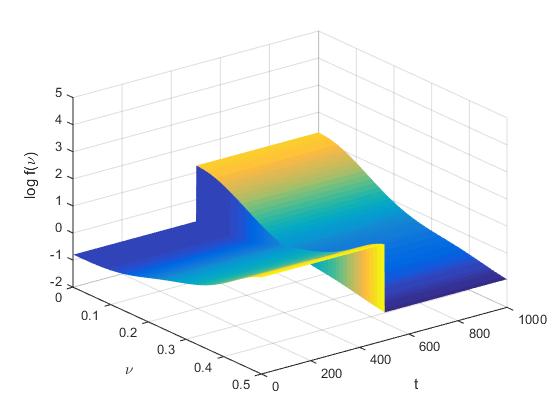}}%
\qquad
\subfloat[]{\includegraphics[height=2in]{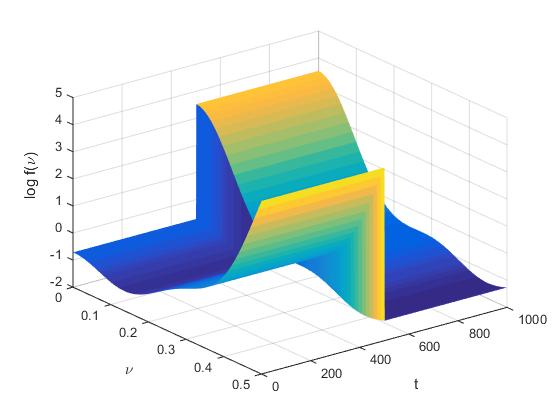}}
\caption{The first row contains the true time-varying spectra for
the covariate modulated piecewise AR process, where $w_{\ell} \le 0.5$ (a)
and $w_{\ell} > 0.5$ (b).  Plots (c) and (d) contain the corresponding
estimated time-varying spectra conditional on two covariate segments.}
\label{fig:simabrupttruefitted}
\end{figure}

The covariate-indexed time-varying spectrum is a 3-dimensional hypersurface, which presents challenges for visualization.  However, when an analysis of the partition points indicates that the process is stationary within distinct covariate segments, the spectrum can be plotted as time-frequency surfaces for each covariate segment. Figure \ref{fig:simabrupttruefitted} displays the true and estimated time-varying log spectrum conditional on two covariate segments.  Both the partition parameter estimates and the log spectra estimates shown here closely approximate the true parameter values and log spectra for this piecewise AR process.

\subsection{Slowly Varying AR Process}
In this section, we consider a collection of $L=8$ slowly-varying AR processes of length $T=1000$ that are modulated by covariates $w_{\ell} = (\ell-1)/(L-1)$ for $\ell=1,\ldots,L$ where
\begin{equation*}
x_{\ell t} = \phi_{\ell t} x_{\ell t-1} + \epsilon_{\ell t},
\label{eqn:smoothreal}
\end{equation*}
 \begin{equation*}
 \phi_{\ell t} =
 \begin{cases}
 -0.5 + t/1000 \quad  \; \;  \mathrm{for} \; 0 \le w_{\ell} \le 0.5 \\
 -0.9 + 9t/5000 \qquad \mathrm{for} \; 0.5 < w_{\ell} \le 1, \\
 \end{cases}
 \label{eqn:smoothcov}
 \end{equation*}
and $\epsilon_{\ell t} \stackrel{\mathrm{iid}}{\sim} N(0,1)$.
In this example, for a given covariate value, spectra change smoothly over time.  

\begin{figure}[t!]%
\centering
\subfloat[]{\includegraphics[height=2in]{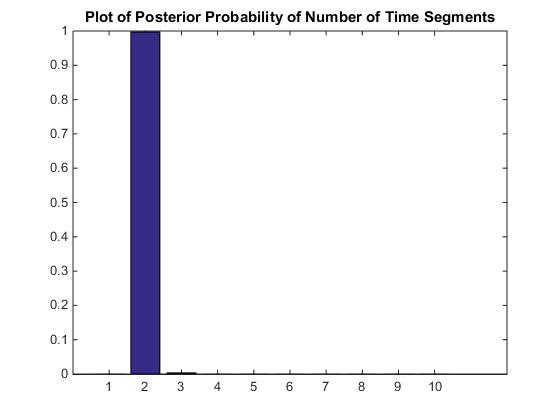}}%
\qquad
\subfloat[]{\includegraphics[height=2in]{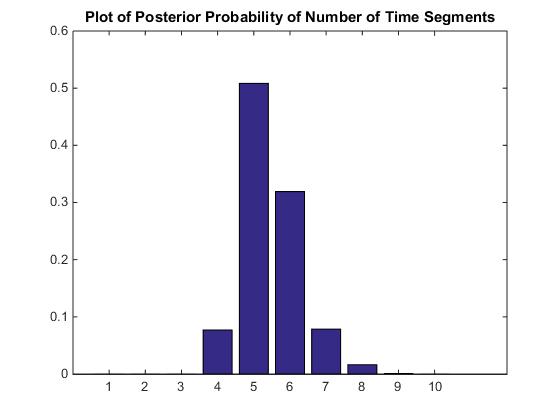}} \\
\subfloat[]{\includegraphics[height=2in]{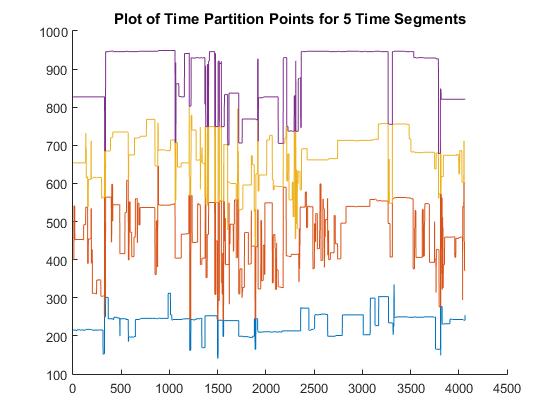}}%
\qquad
\subfloat[]{\includegraphics[height=2in]{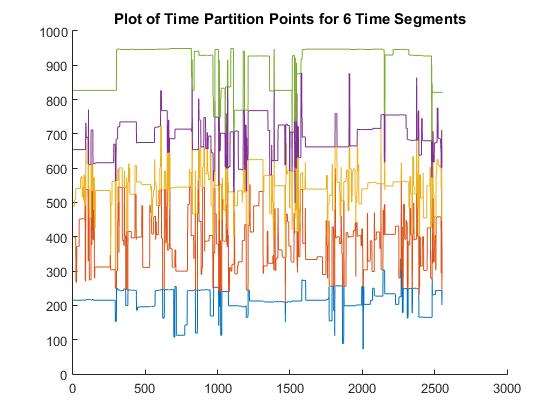}} 
\caption{Posterior probabilities for the number of time segments for
  the piecewise AR process (a) and the slowly-varying AR process (b) and time-varying distribution of time partition points for iterations with five (c) and six (d) time segments for the slowly-varying AR process.}
\label{fig:simposttime}
\end{figure}
\begin{figure}[t!]%
\centering
\subfloat[]{\includegraphics[height=2in]{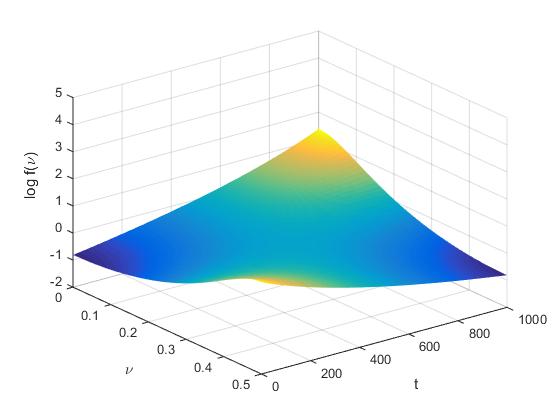}}%
\qquad
\subfloat[]{\includegraphics[height=2in]{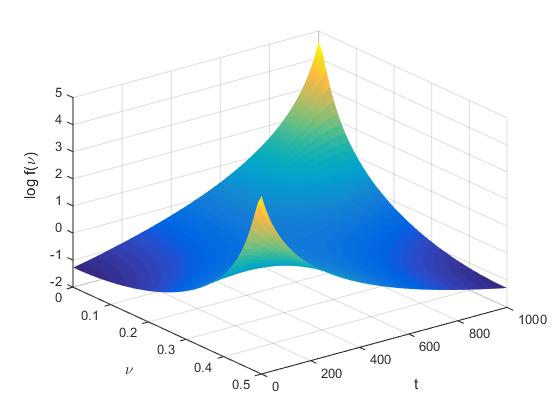}} \\
\subfloat[]{\includegraphics[height=2in]{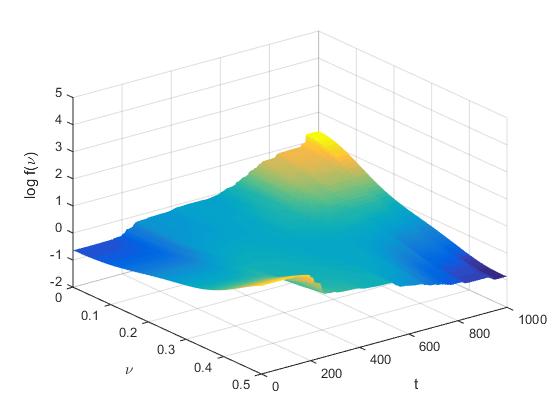}}%
\qquad
\subfloat[]{\includegraphics[height=2in]{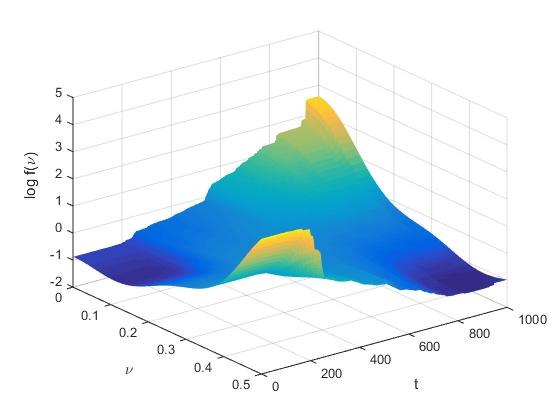}}
\caption{The first row contains the true time-varying spectra for
the covariate-modulated slowly-varying  AR process, where (a) $w \le -0.5$
and (b) $w > 0.5$.  Plots (c) and (d) contain the corresponding
estimated time-varying spectra conditional on two covariate segments.}
\label{fig:simsmoothtruefitted}
\end{figure}

The top row of Figure \ref{fig:simposttime} contains the posterior distribution for the number of time segments for the slowly-varying AR process.  Note that more time segments are proposed compared to the piecewise AR process as we are using a piecewise model approximation to the slowly varying process.  Additionally, the bottom row of Figure \ref{fig:simposttime} shows how the distribution of time partition points changes from one iteration to the next for iterations with five and six time segments.  Our proposed approach can recover slowly varying processes because the spectral estimator averages over the possible locations of the time partition points reflected in the posterior distribution.  By contrast, the distribution of the time partition point for the piecewise AR process is tightly packed around the true partition point, so we are averaging across a much smaller range of possible partition points.  The model again estimates two covariate segments with a covariate partition point at the largest realized value less than 0.5, and Figure \ref{fig:simsmoothtruefitted} displays the true and estimated time-varying log spectra conditional on the two covariate segments.

\section{Application to AgeWise Caregiver Study}
\label{sec:application}
We now present analytical results of applying the proposed method to the motivating study described in Section \ref{sec:motivatingstudy}, which consists of HRV data for the 5 minutes before and 5 minutes after entering the first period of REM sleep and associated PSQI scores from $L=30$ study participants.

\subsection{Time-Varying Spectra}
The proposed procedure identified one clear temporal partition with the probability of $m=2$ estimated as 99.99\%.  The one partition point occurs 2.3 minutes before the onset of REM with a probability of 99.80\%.  In our study, sleep staging was determined by trained technicians who inspected the electroencephalogram (EEG) following established guidelines \citep{iber2007}.   Consequently, our results indicate that changes in autonomic nervous system activity, as measured by HRV, precede neurological changes, as measured by EEG.  This finding is not completely unexpected, as preliminary results have been reported in which modifications in cardiac vagal activity precede changes in EEG power spectrum \citep{brainheartlink}.

\begin{figure}[t!]%
\centering
\subfloat{\includegraphics[height=2in]{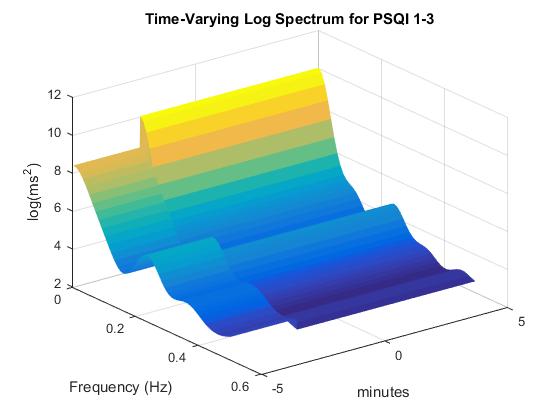}}%
\qquad
\subfloat{\includegraphics[height=2in]{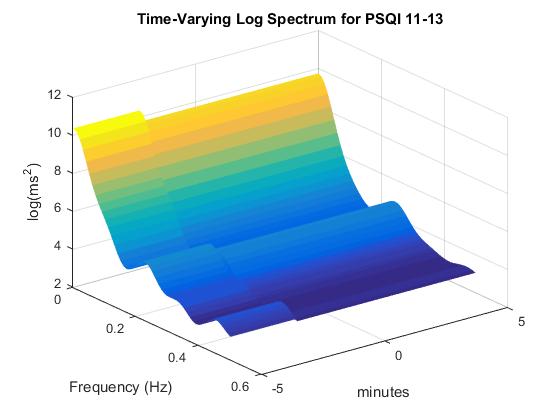}} 
\caption{Estimated time-varying log spectra for the 5 minutes before and 5 minutes after the first period of rapid eye movement sleep for PSQI scores 1-3 and 11-13 respectively.}
\label{fig:applogspectra}
\end{figure}

With regards to PSQI, the procedure separated the data into 4 segments:  PSQI from 1--3, 4--6, 7--10 and 11--13.  Recalling that PSQI greater than 6 is typically indicative of clinically disturbed sleep, we refer to these groups as excellent, good, fair and poor sleep quality, respectively.   Figure \ref{fig:applogspectra} presents the estimated time-varying log spectra for subjects in the two extreme groups:  excellent and  poor sleep quality.  There are two noticeable differences between these two time-varying spectra.  First, power in low frequencies (LF) from 0.04-0.15 Hz increases between NREM and REM for participants with excellent sleep quality, but decreases for poor sleep quality.  Second, the change in power from NREM to REM for high frequencies (HF) between 0.15-0.40 Hz is less drastic for participants with poor sleep quality than it is for those with excellent sleep quality.  

Relative power within these HF and LF bands is used by researchers as measures of autonomic nervous system activity, which is believed to play an important role in the rejuvenating properties of sleep \citep{Vanoli1918}.  Consequently, differences in the dynamics of HF and LF power between people with excellent and poor sleep quality may explain how poor sleep is associated with ill-health.  A favorable property of the proposed method is that the RJMCMC provides a means to estimate and conduct inference on any function of the covariate-indexed time-varying spectrum.  In the following subsection, we use this ability to conduct inference on the association between PSQI and these collapsed measures.


\subsection{Collapsed Measures}
The parasympathetic branch of the autonomic nervous system is responsible for bodily activities that occur while at rest and its modulation  is inversely related to stress and arousal.  The amount of HF power relative to the amount of combined power from HF and LF bands, which is referred to as normalized-HF (HFnu), provides a measure of parasympathetic modulation \citep{HRVstandards}.  From the covariate-indexed time-varying spectrum, we can define HFnu as a function of PSQI and time 
\[ {\tt HFnu}(u,w) = \left[\int_{0.15}^{0.40} f(u, w, \nu) d\nu\right] \bigg/ \left[\int_{0.04}^{0.40} f(u, w, \nu) d\nu\right]. \]

As previously mentioned, the method automatically provides point estimates and credible intervals for any function of the conditional power spectrum, including 
HFnu, through the mean and percentiles of the sample generated from the MCMC algorithm.  Estimated ${\tt HFnu}(u,w)$ within each of the time segments for each of the covariate segments provides a closer exploration of this dynamic relationship.  For those with excellent sleep quality (PSQI 1--3), the estimated HFnu and 95\% pointwise credible intervals in NREM and REM are 0.27 (0.21, 0.34) and 0.21 (0.17, 0.24), respectively.  For those with poor sleep quality (PSQI 11--13), the estimated HFnu and 95\% pointwise credible intervals in NREM and REM are 0.09 (0.07, 0.11) and 0.15 (0.13, 0.17), respectively.  

These intervals show that HFnu is significantly higher across the entire period of observation for those with excellent sleep quality as compared to those with poor.   This finding indicates that people with poor sleep quality do not exhibit the expected fluctuations in autonomic nervous system arousal that are seen during sleep in individuals with excellent sleep quality.  These intervals also show that the decrease in HFnu from NREM to REM in participants with excellent sleep quality and the increase in those with poor sleep quality are statistically significant.  In healthy individuals, one expects parasympathetic activity to decrease during REM, in which dreaming, eye movement and body movement commonly occur \citep{Vanoli1918}. Our results suggest that not only do individuals with poor sleep quality exhibit increased autonomic nervous system arousal during sleep, but the dynamics of their parasympathetic activity across sleep periods differs from good sleepers, representing two potential pathways through which poor sleep quality may be linked to ill-health and functioning.

\section{Discussion}
\label{sec:discussion}
This article proposes a method for simultaneous and automatic analysis of the association between the time-varying spectrum and study covariates.  A locally stationary model indexed by time and covariate is formulated with smooth estimation of the local spectra.  The MCMC estimation procedure allows the time and covariate indices to vary from one iteration to the next for adaptive estimation of the time and covariate-varying spectrum.  The Bayesian formulation and estimation procedure also allow for the estimation of abrupt or smooth changes in the spectrum by averaging over MCMC iterations as shown in the simulation setting.  This approach is motivated by and used to analyze the association between the time-varying spectrum of heart rate variability and self-reported sleep quality in a population of older adults who are the primary caregiver for their ill spouse.  The analysis provides insight on the dynamic relationship between autonomic arousal, a physiological marker of stress, and sleep, which can serve as a guide for designing behavioral interventions to enhance the lives of caregivers.

This article is one of the first approaches to analyzing the power spectrum of replicated nonstationary time series indexed by a set of covariates and is not meant to be plenary.  The power spectra of replications with similar covariate values may exhibit some variability \citep{krafty2011,withingroupvariability}, so a topic of future research would be to incorporate within-group spectral variability into the modeling framework.  Further, the presented approach allows for the analysis of replicated, univariate time series. In many applications, interest lies in how covariates are associated with the power spectrum of multivariate time series.  For instance, the simultaneous spectral analysis of HRV and EEG can provide insights into how the dynamic coupling of the heart and brain are connected to clinical and behavioral outcomes \citep{rothenberger2015}.  An extension of the framework to the multivariate setting is also a topic of future research. 


\backmatter


\section*{Acknowledgements}
This work is supported by NIH grants R01GM113243, P01AG20677 and R01HL104607.





%

\bibliographystyle{biom} \bibliography{mybib}





 \appendix


\section*{Web Supplement A: AgeWise Caregiver Study Data}
\label{sec:websuppA}
Demeaned heart rate variability (HRV) time series for each subject considered in this study are included here.  As mentioned in the article, each participant served as the primary caregiver for their spouse who was suffering from a progressive dementing illness such as Alzheimer’s or advanced Parkinson’s disease. We isolated a 10 minute long epoch of HRV during the 5 minutes before and 5 minutes after the first onset of REM sleep.  Pittsburgh Sleep Quality Index scores are also reported as a self-reported measure of sleep disturbances and of how these disturbances affect daily functioning over a one-month period.


\begin{center}
\includegraphics[width=0.5\textwidth]{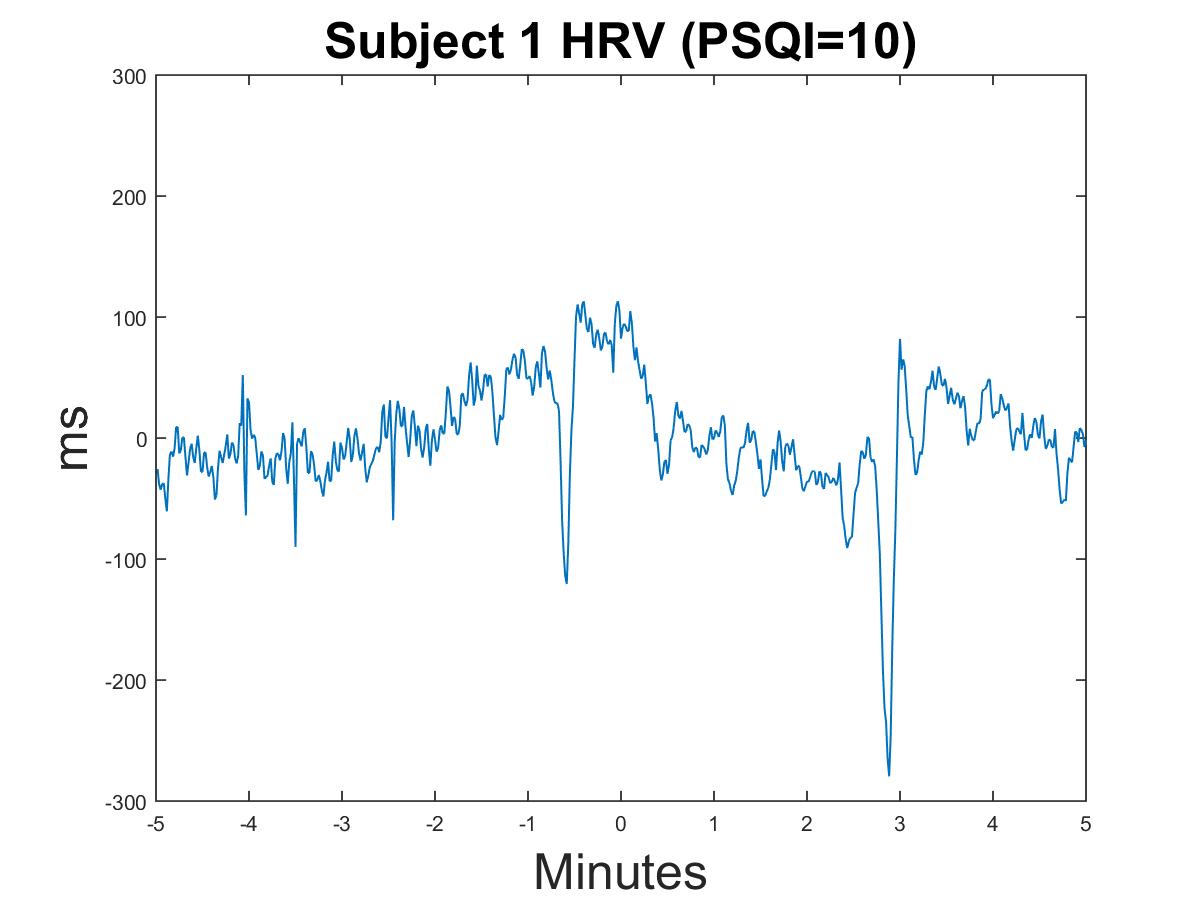}\includegraphics[width=0.5\textwidth]{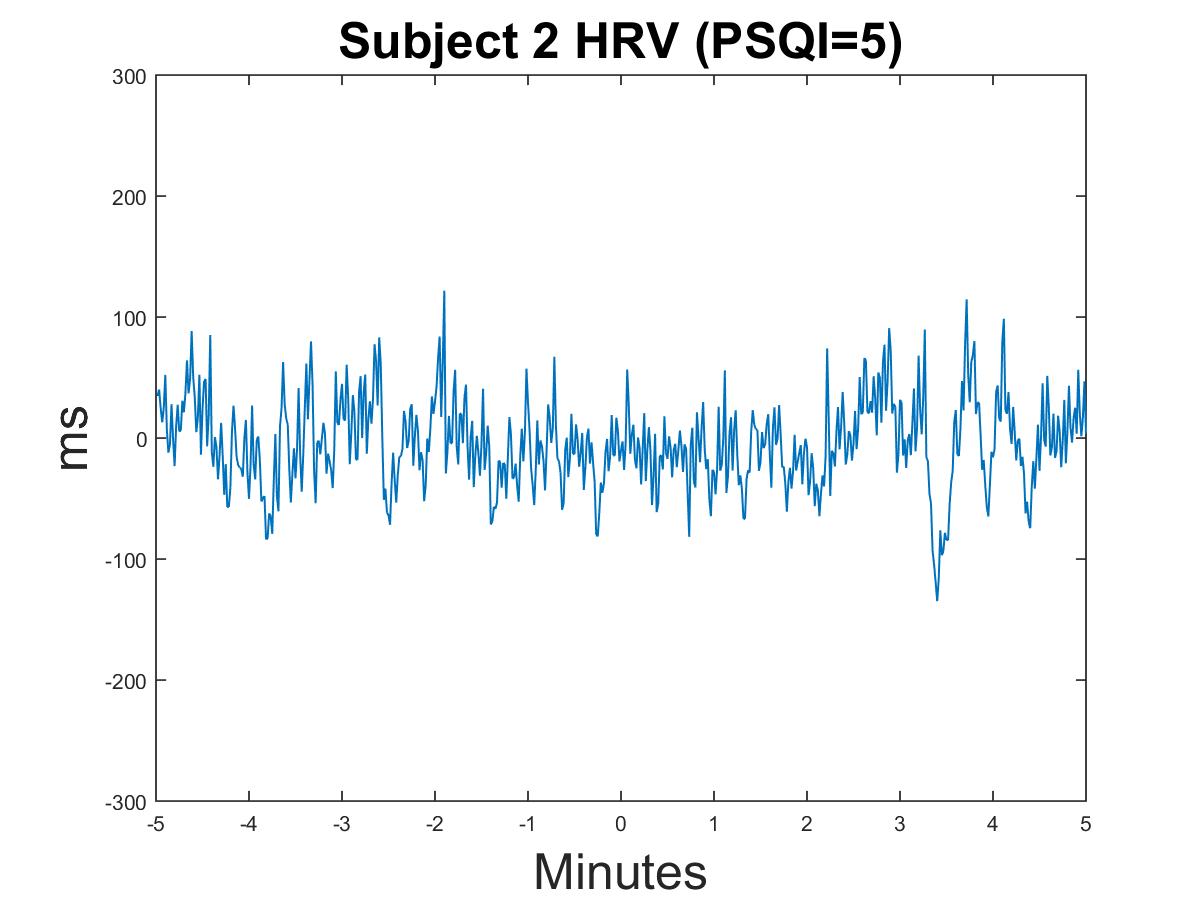}\\
\vspace{2cm}
\includegraphics[width=0.5\textwidth]{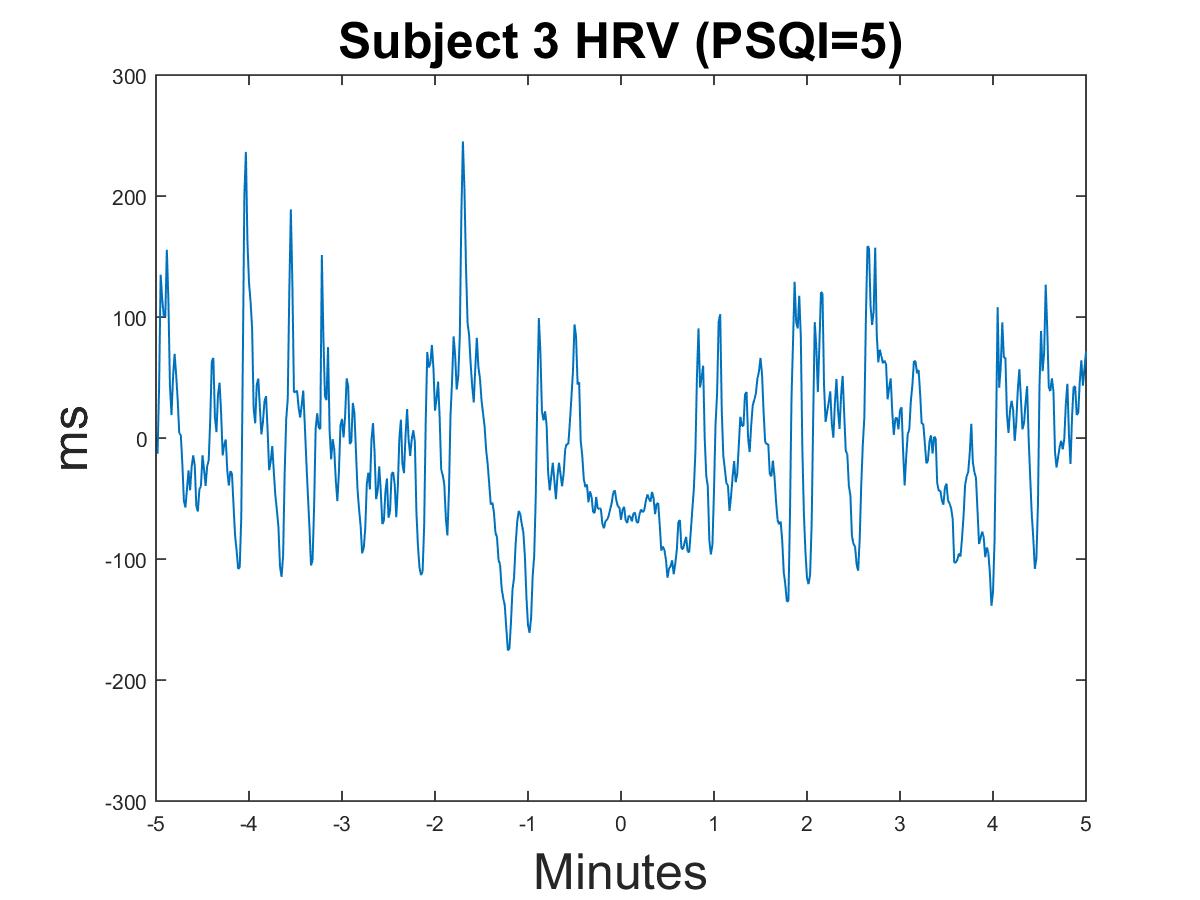}\includegraphics[width=0.5\textwidth]{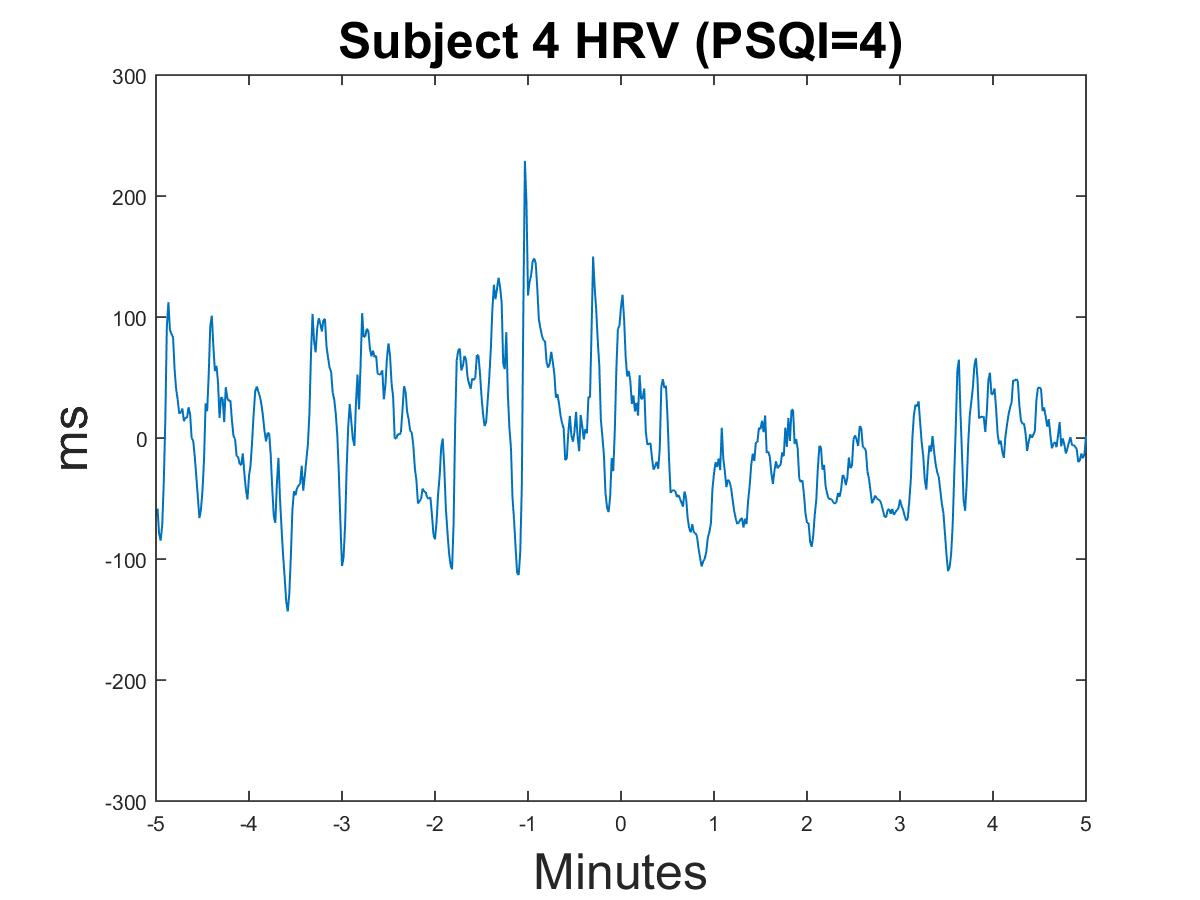}\\

\includegraphics[width=0.5\textwidth]{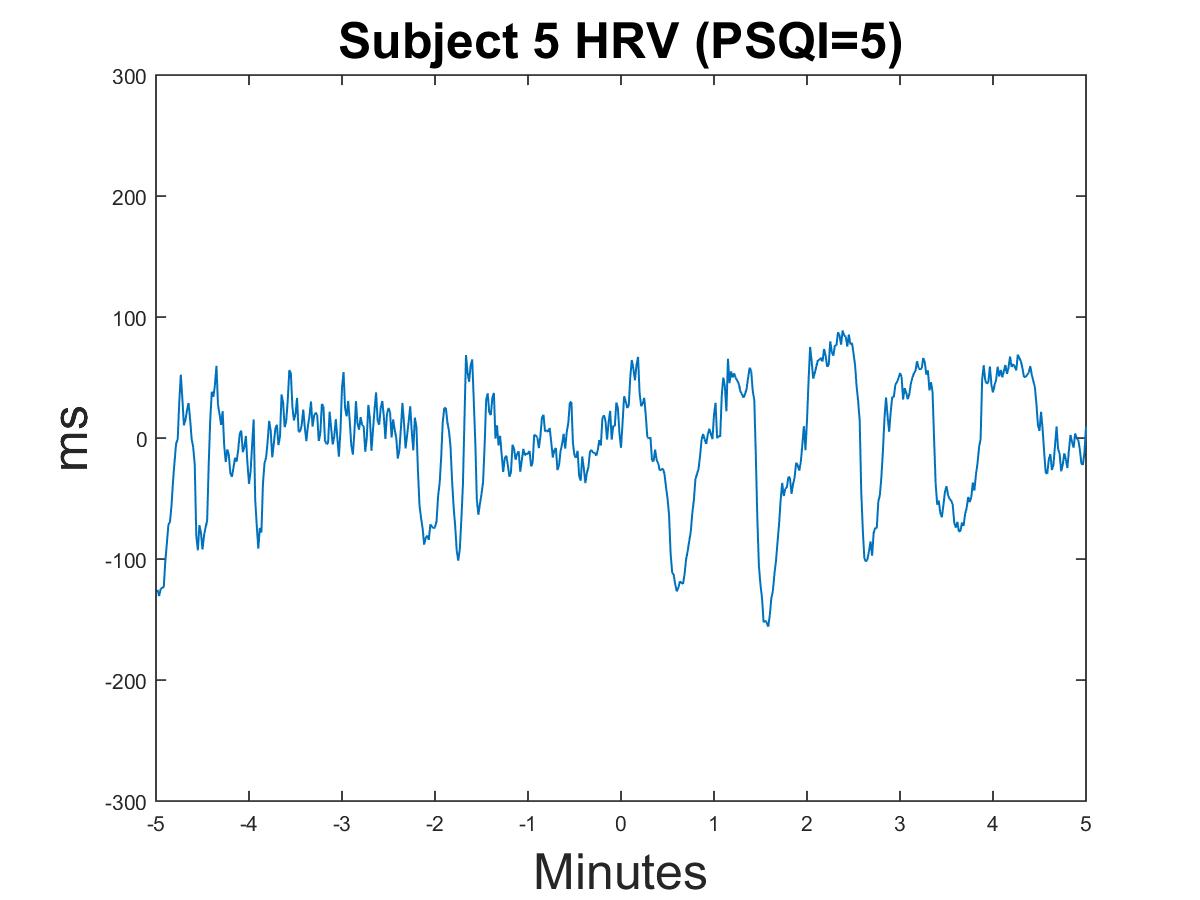}\includegraphics[width=0.5\textwidth]{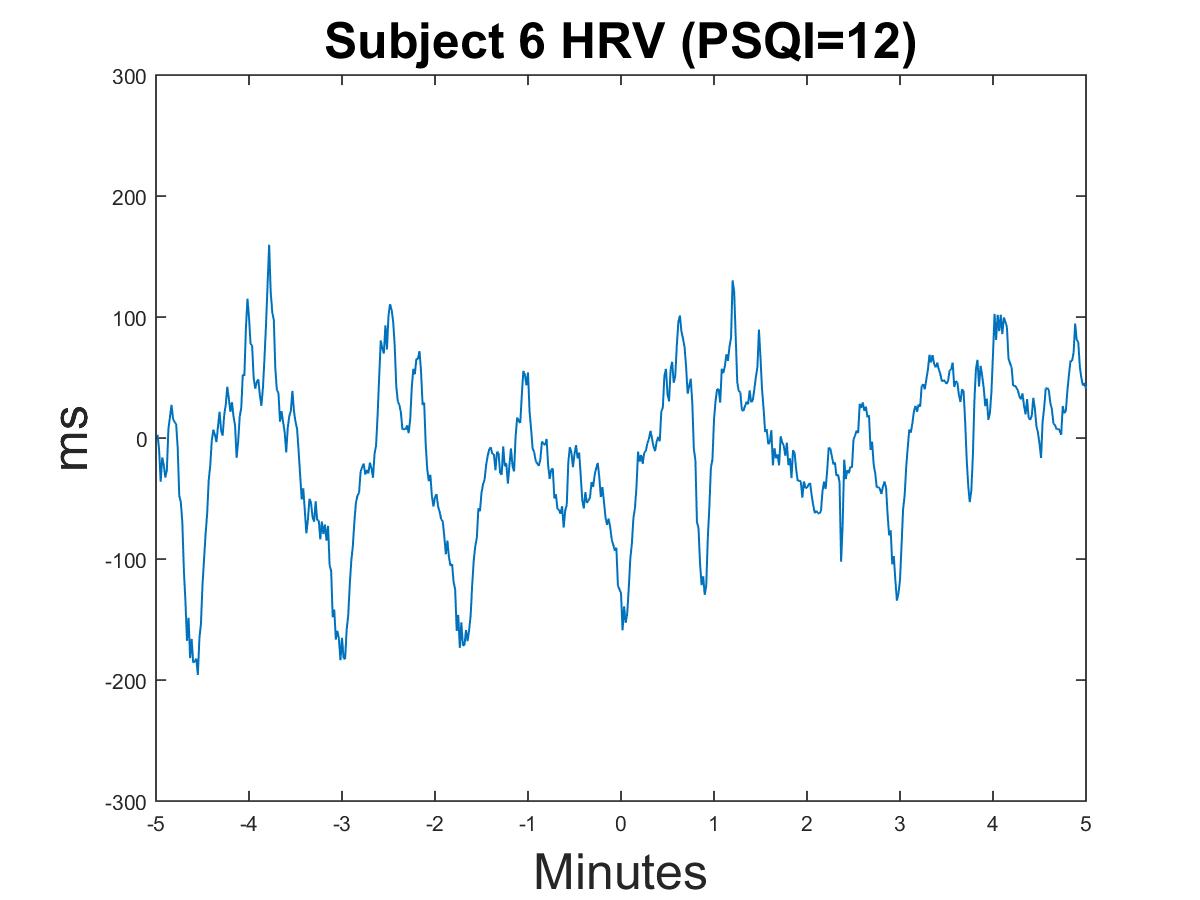}\\
\vspace{2cm}
\includegraphics[width=0.5\textwidth]{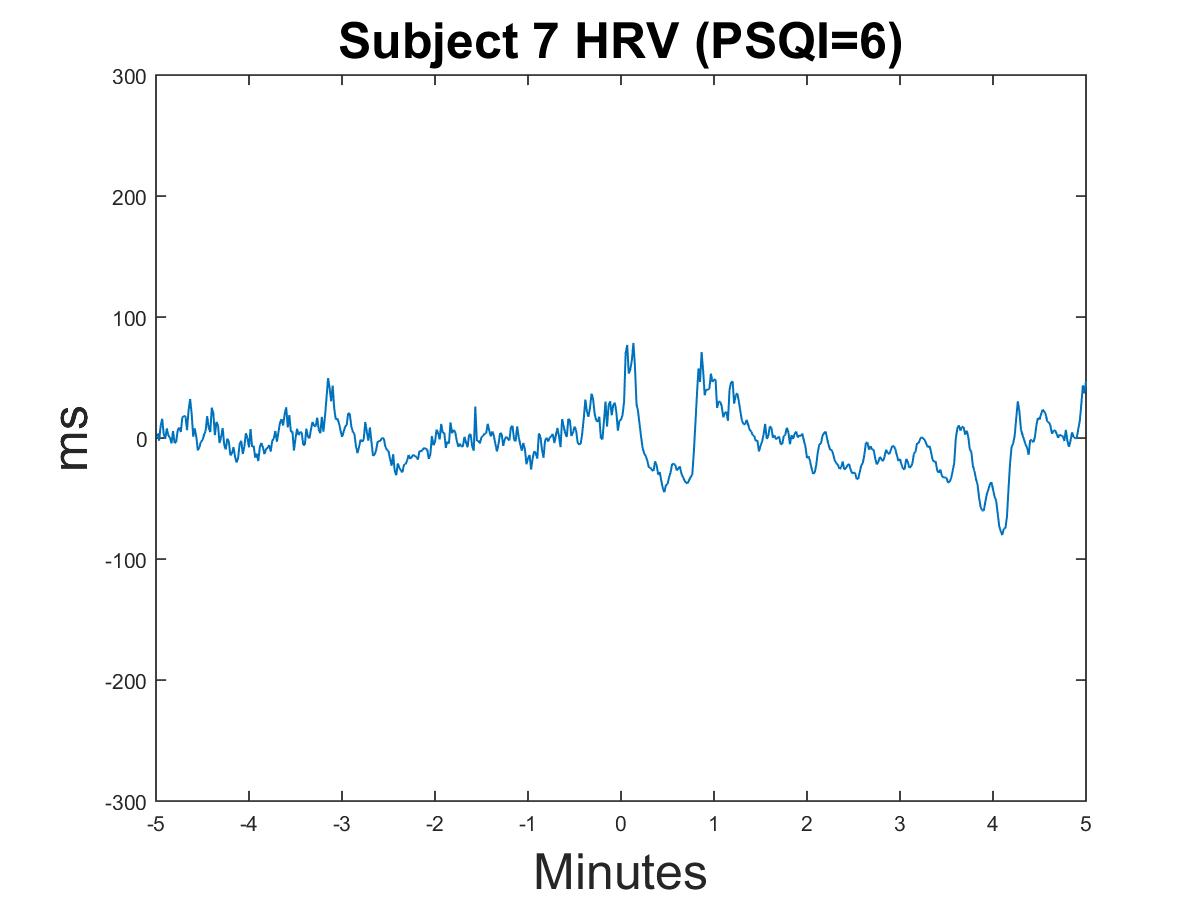}\includegraphics[width=0.5\textwidth]{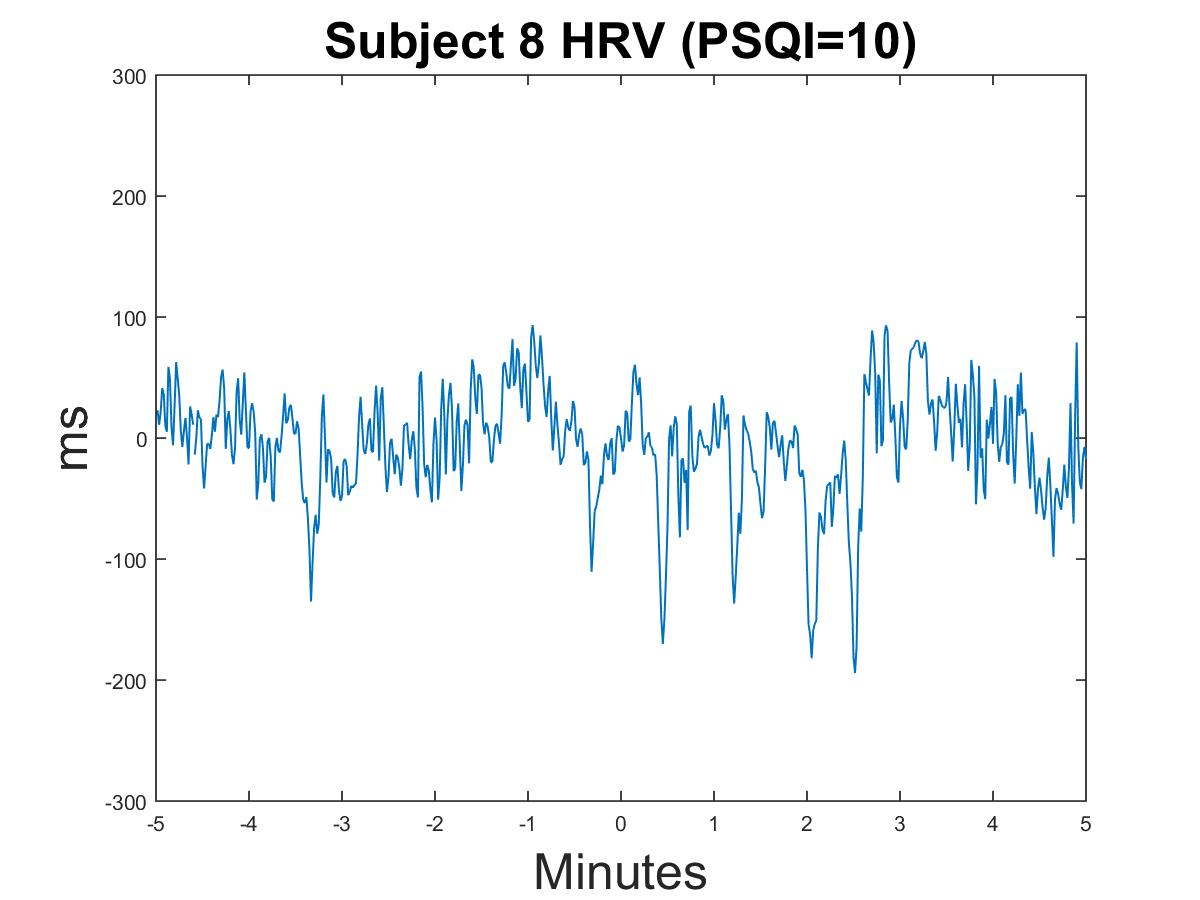}\\
\vspace{2cm}
\includegraphics[width=0.5\textwidth]{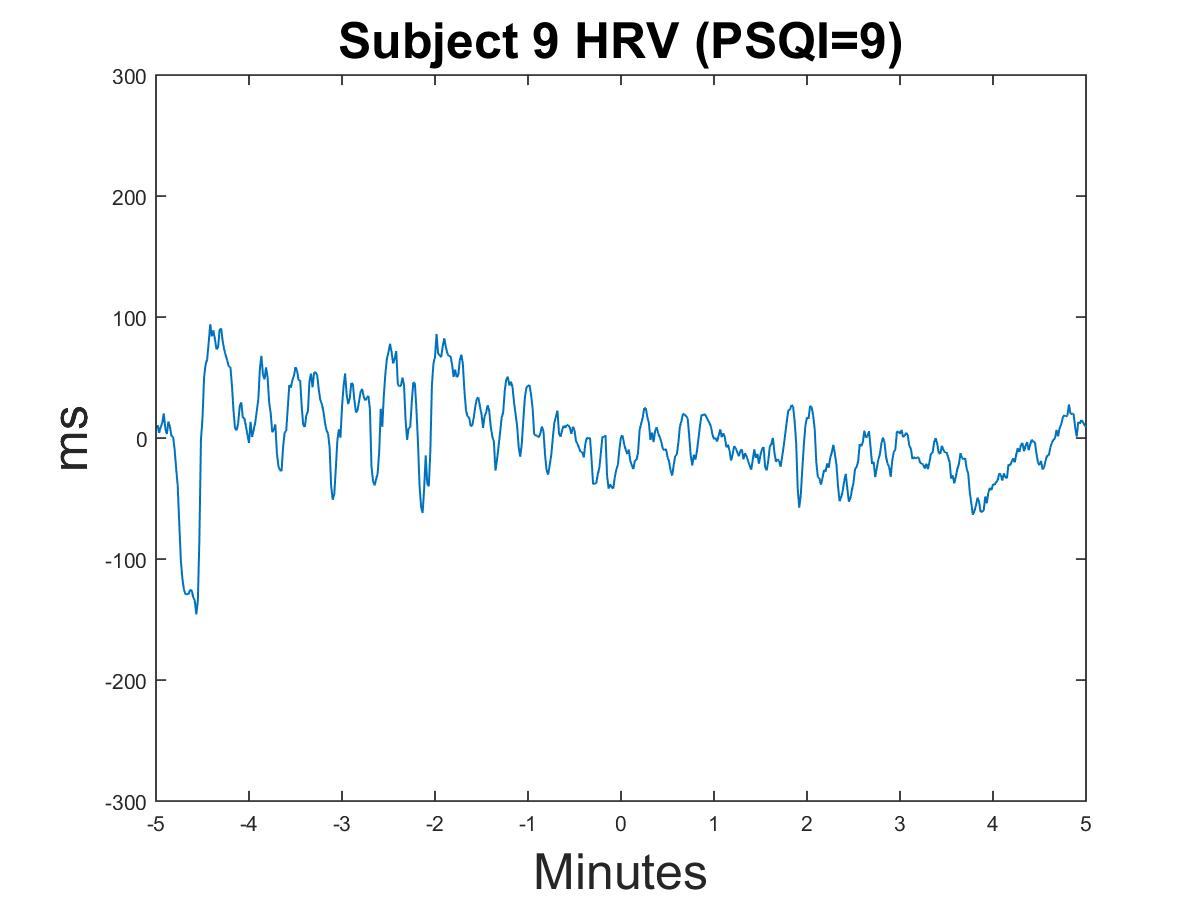}\includegraphics[width=0.5\textwidth]{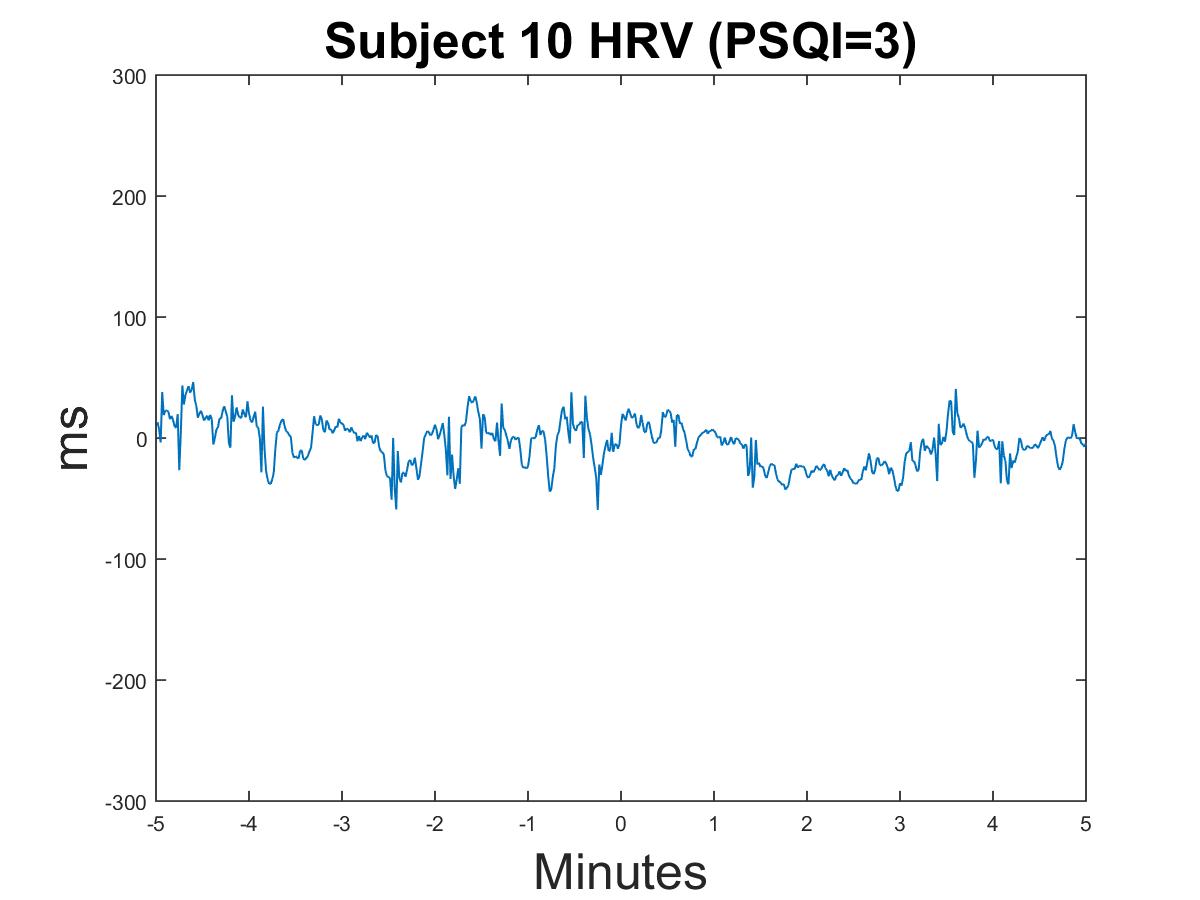}\\

\includegraphics[width=0.5\textwidth]{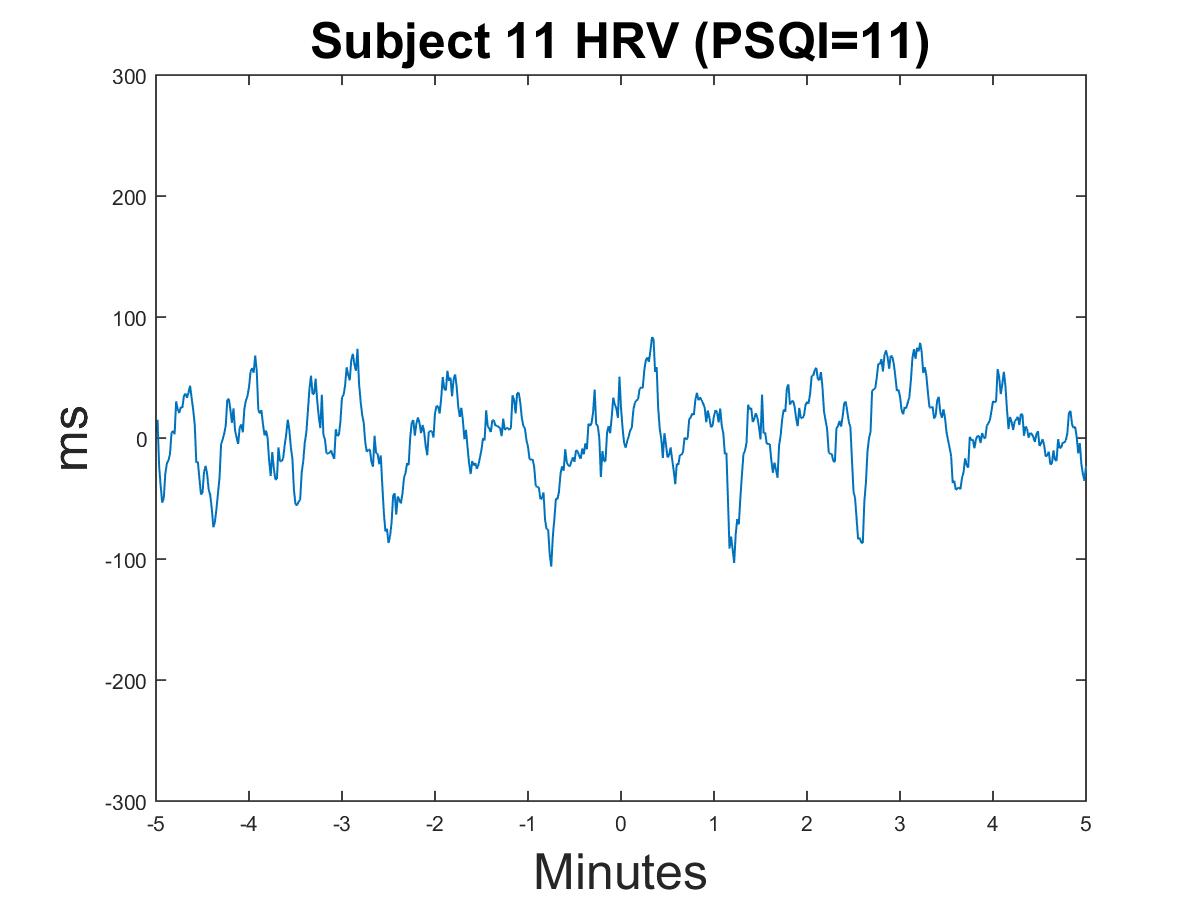}\includegraphics[width=0.5\textwidth]{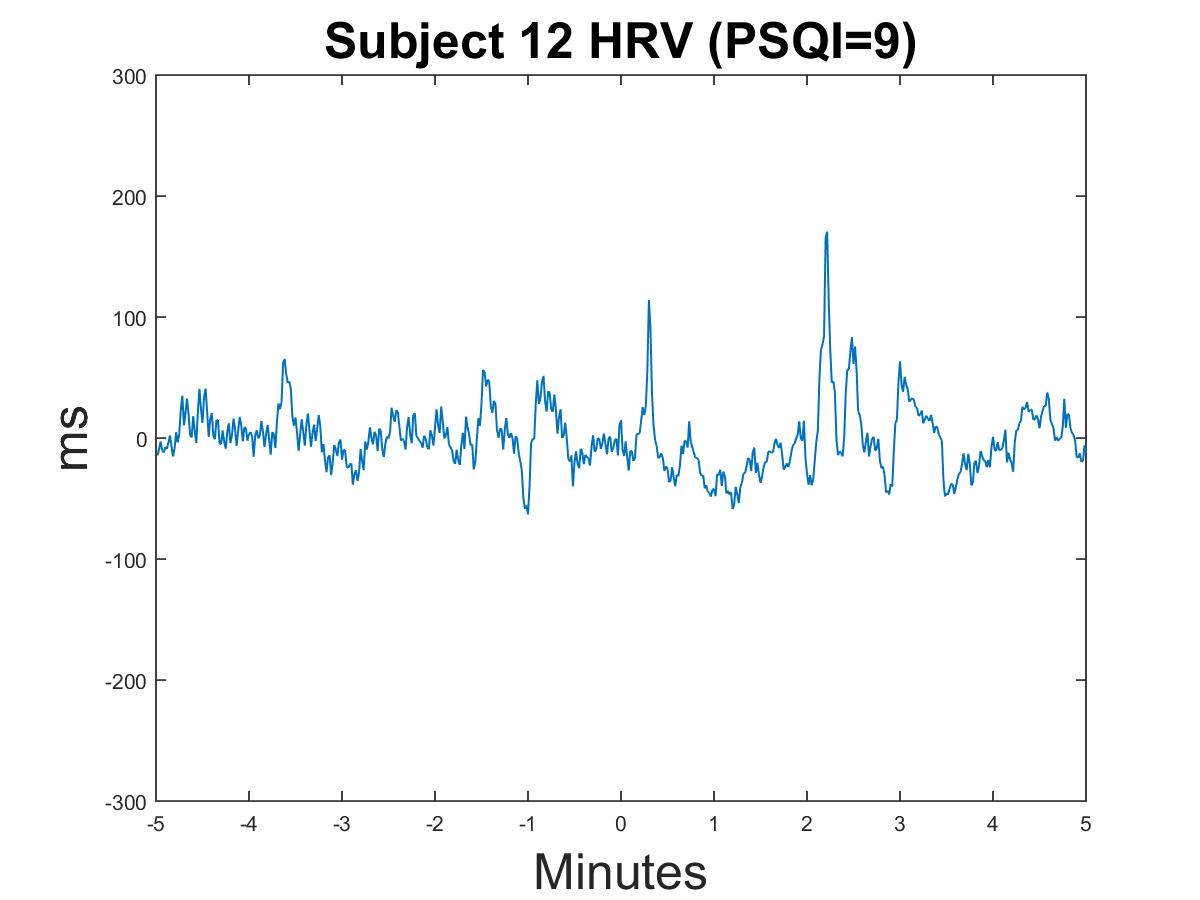}\\
\vspace{2cm}
\includegraphics[width=0.5\textwidth]{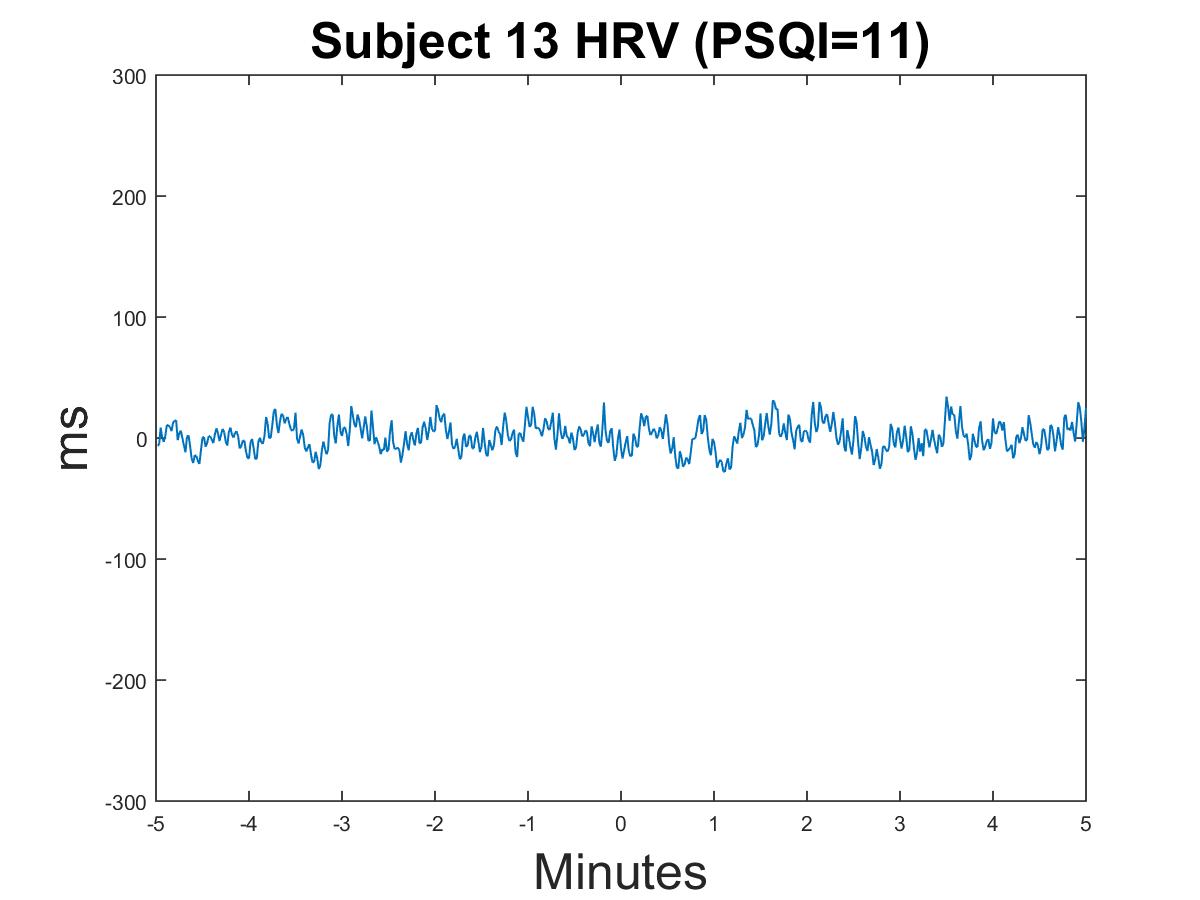}\includegraphics[width=0.5\textwidth]{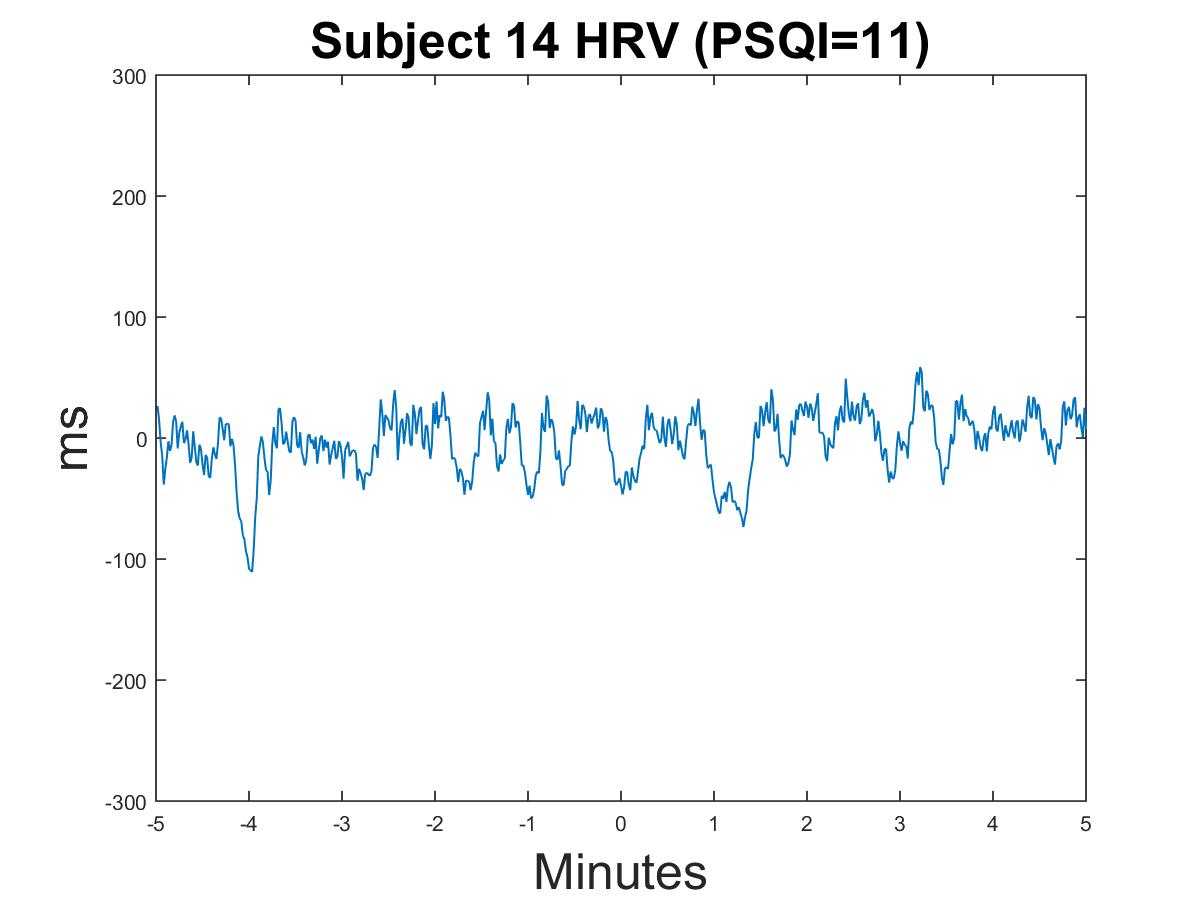}\\
\vspace{2cm}
\includegraphics[width=0.5\textwidth]{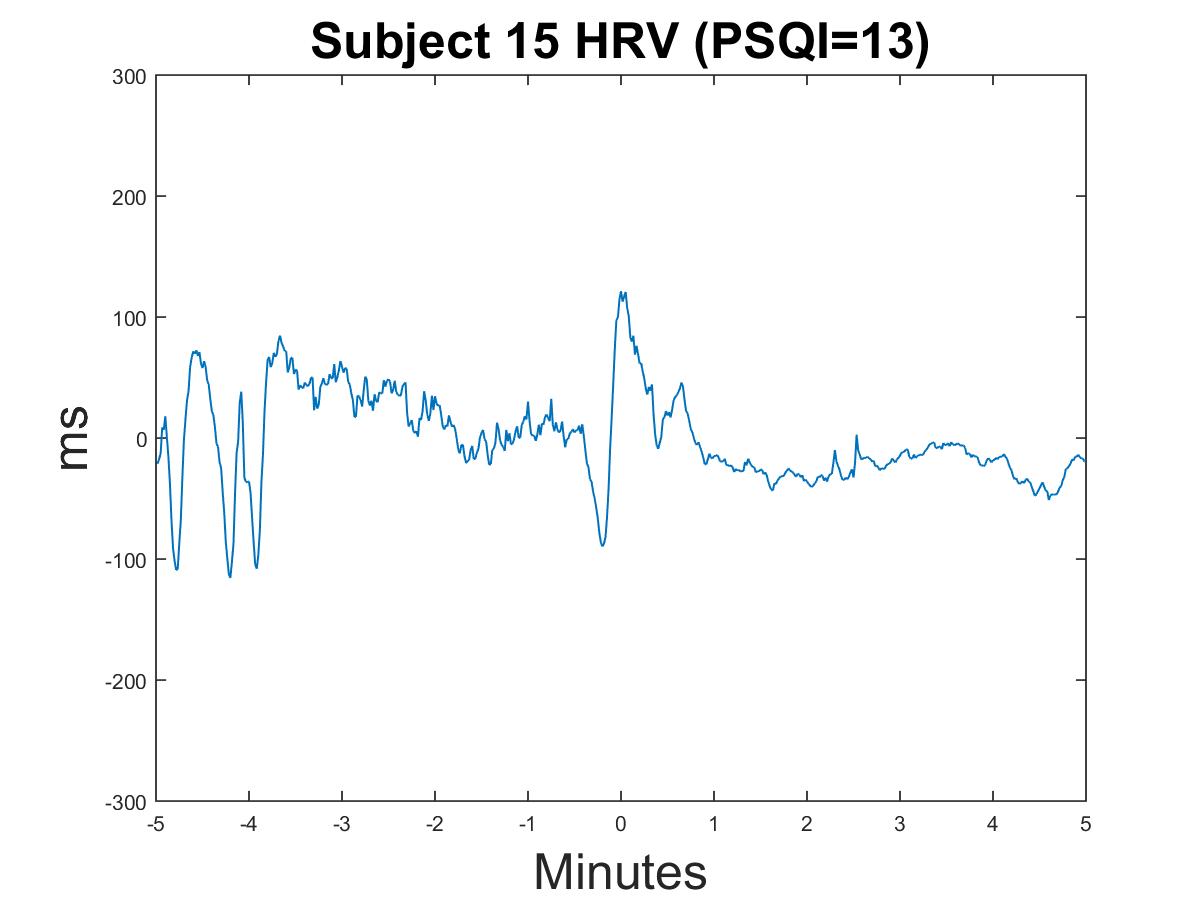}\includegraphics[width=0.5\textwidth]{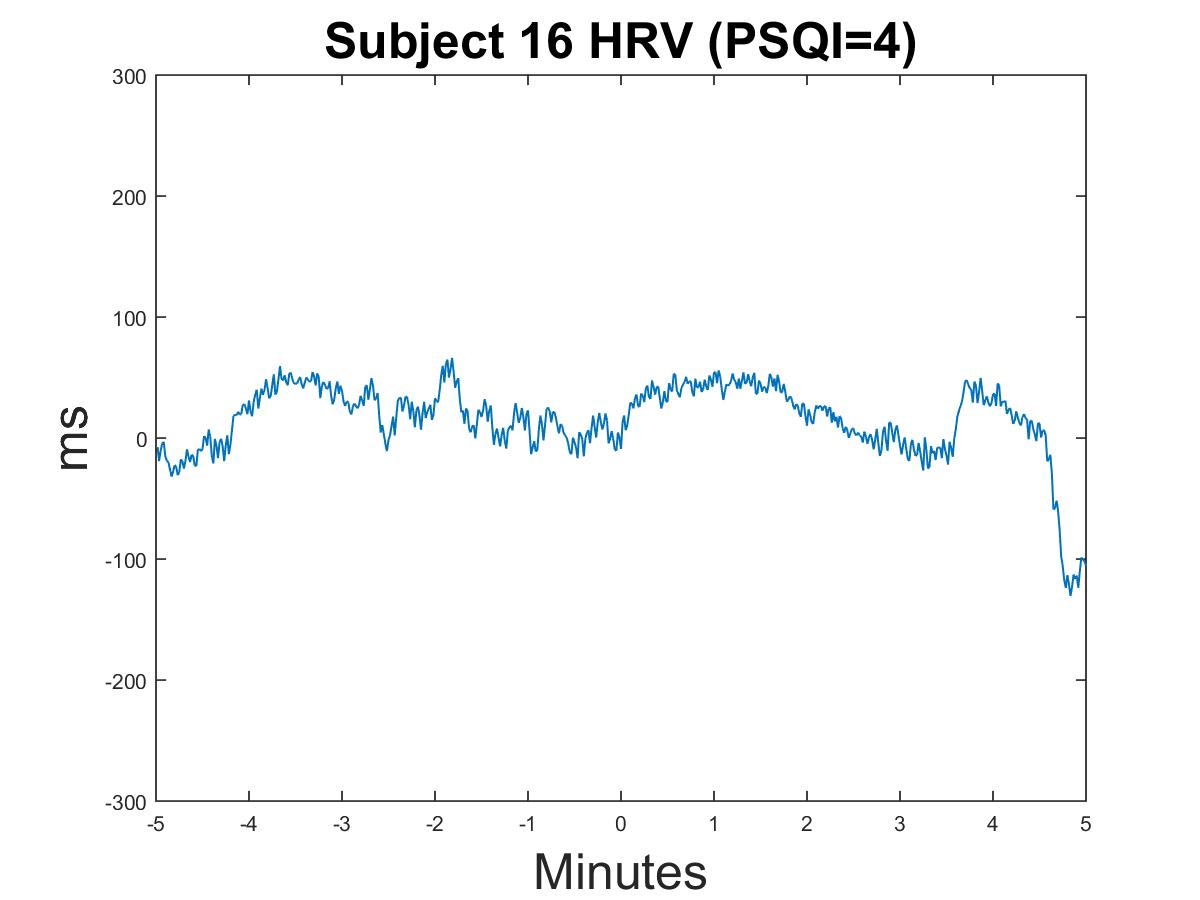}\\

\includegraphics[width=0.5\textwidth]{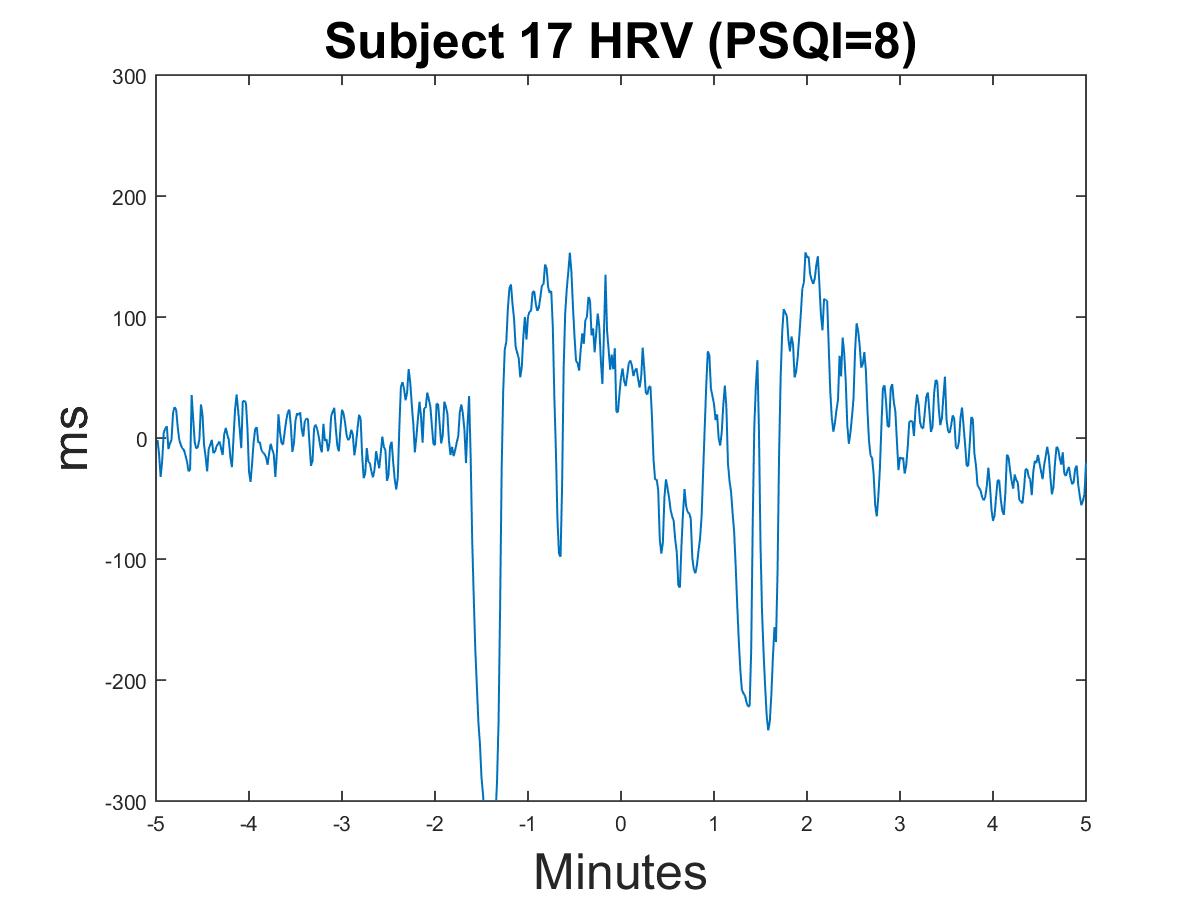}\includegraphics[width=0.5\textwidth]{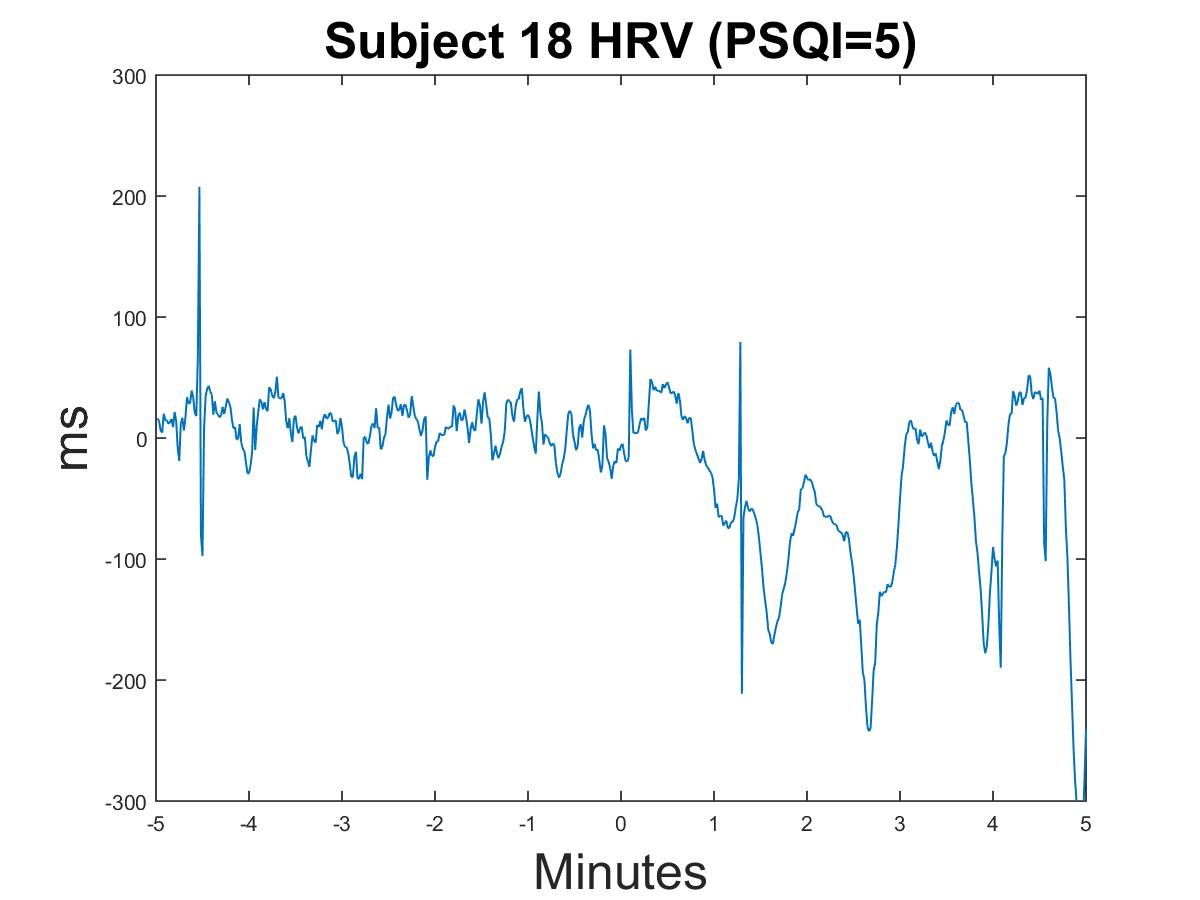}\\
\vspace{2cm}
\includegraphics[width=0.5\textwidth]{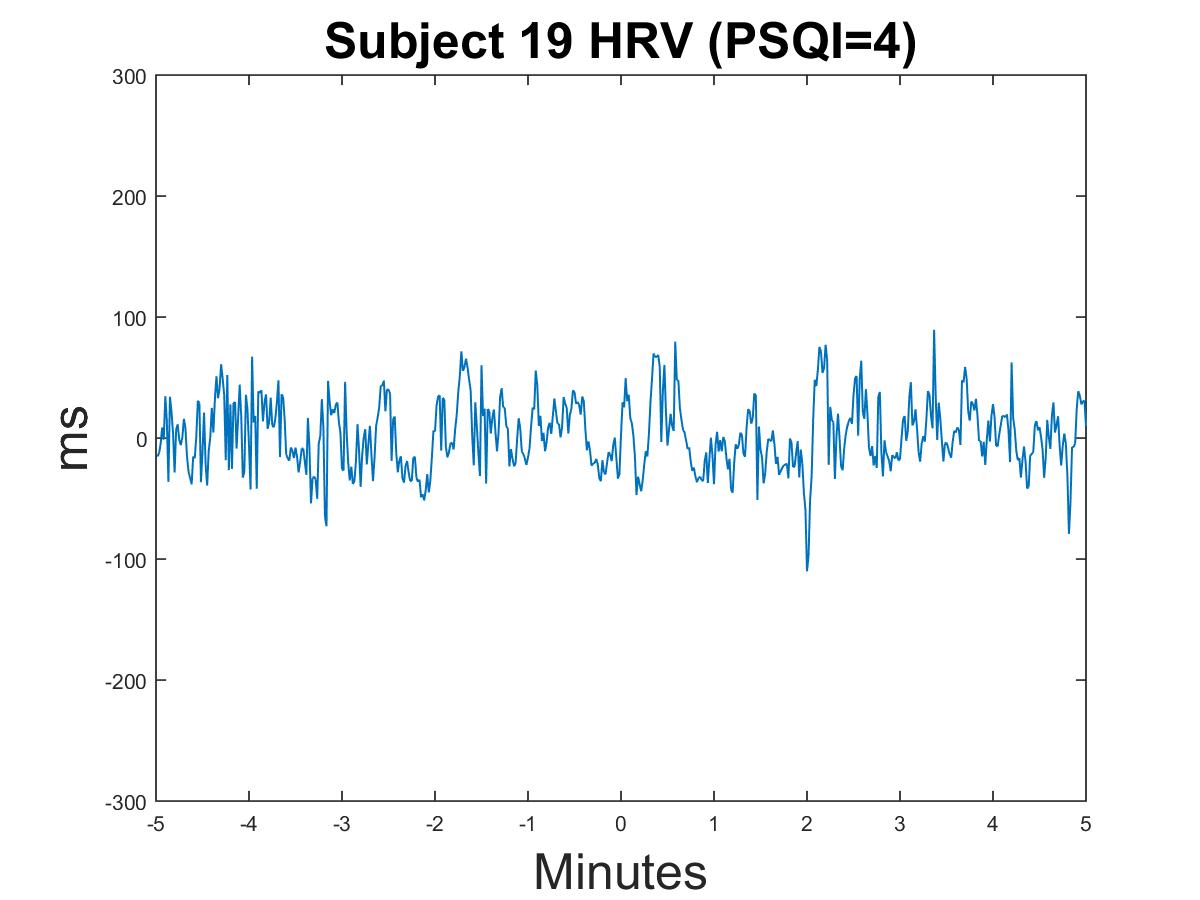}\includegraphics[width=0.5\textwidth]{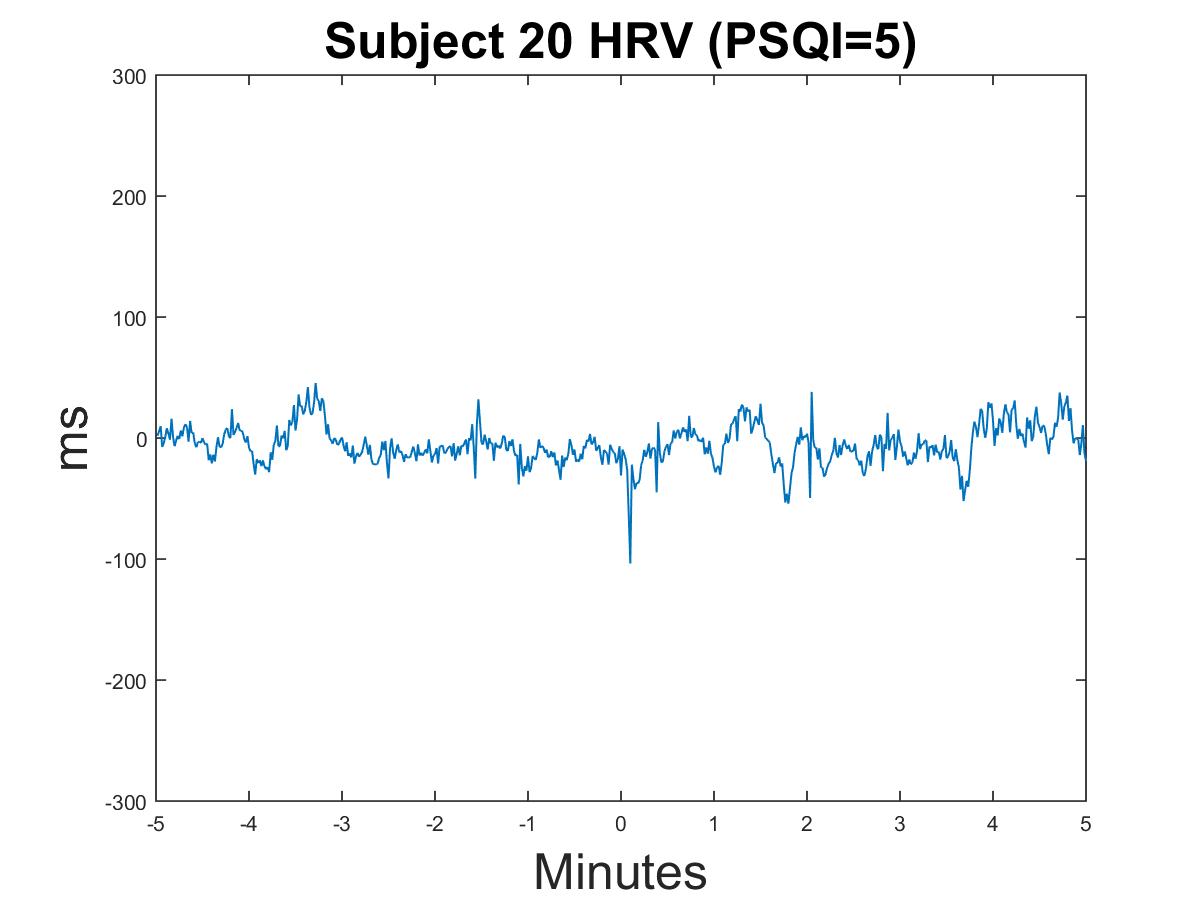}\\
\vspace{2cm}
\includegraphics[width=0.5\textwidth]{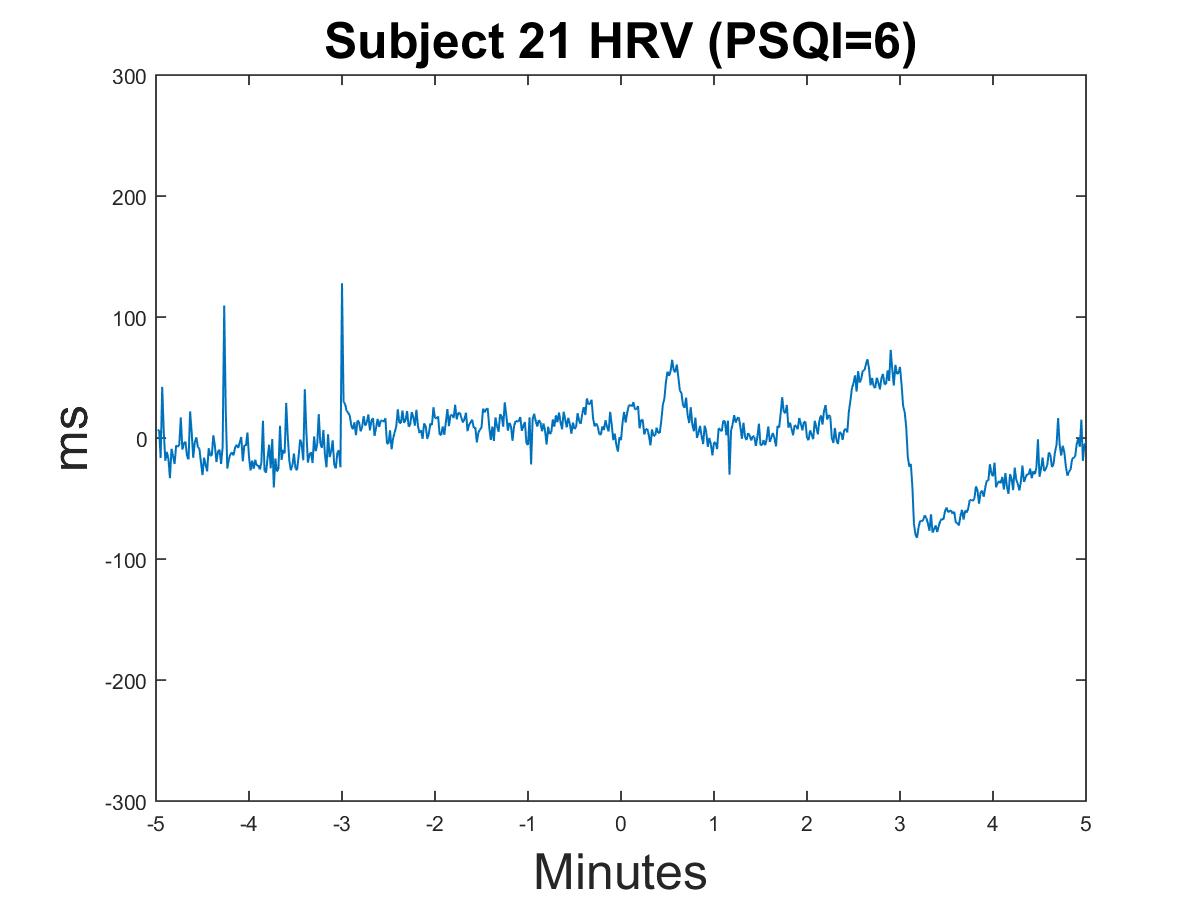}\includegraphics[width=0.5\textwidth]{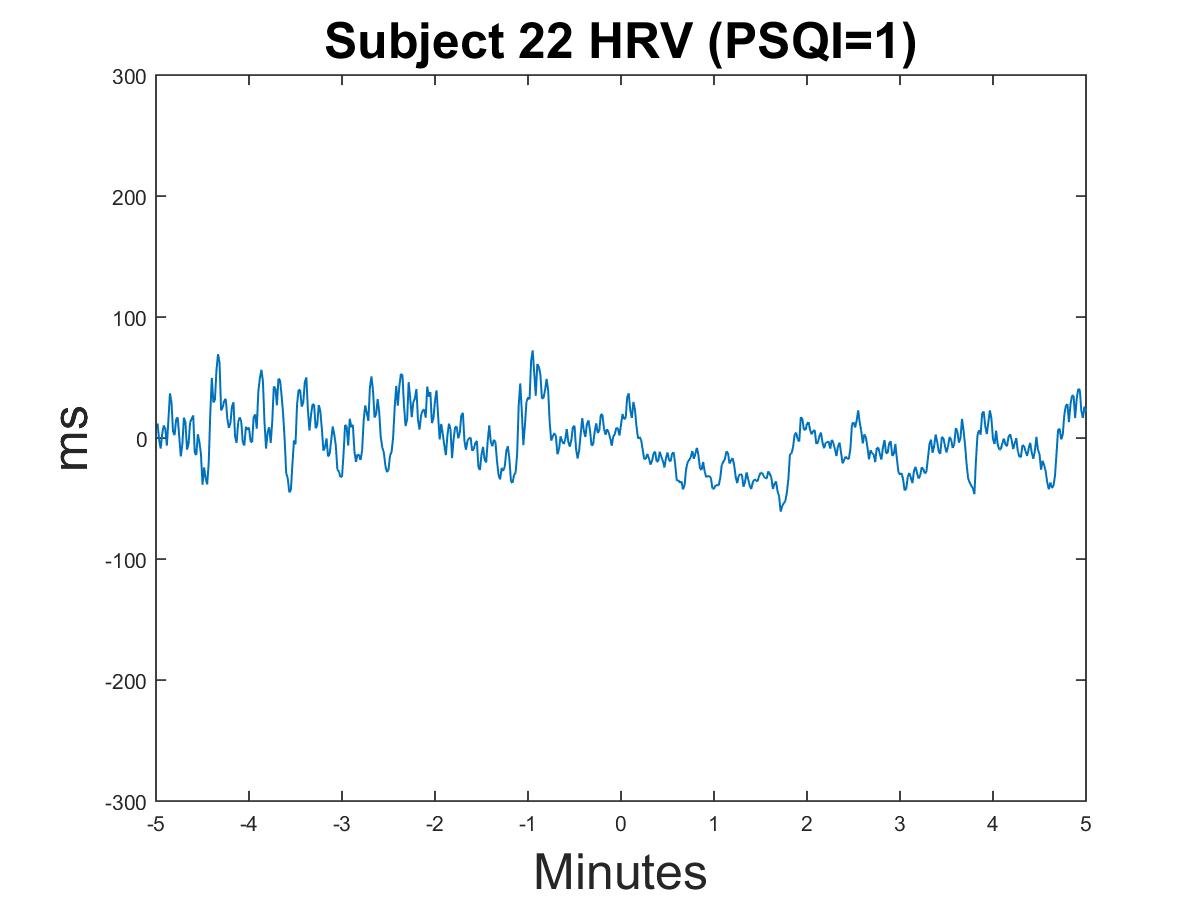}\\

\includegraphics[width=0.5\textwidth]{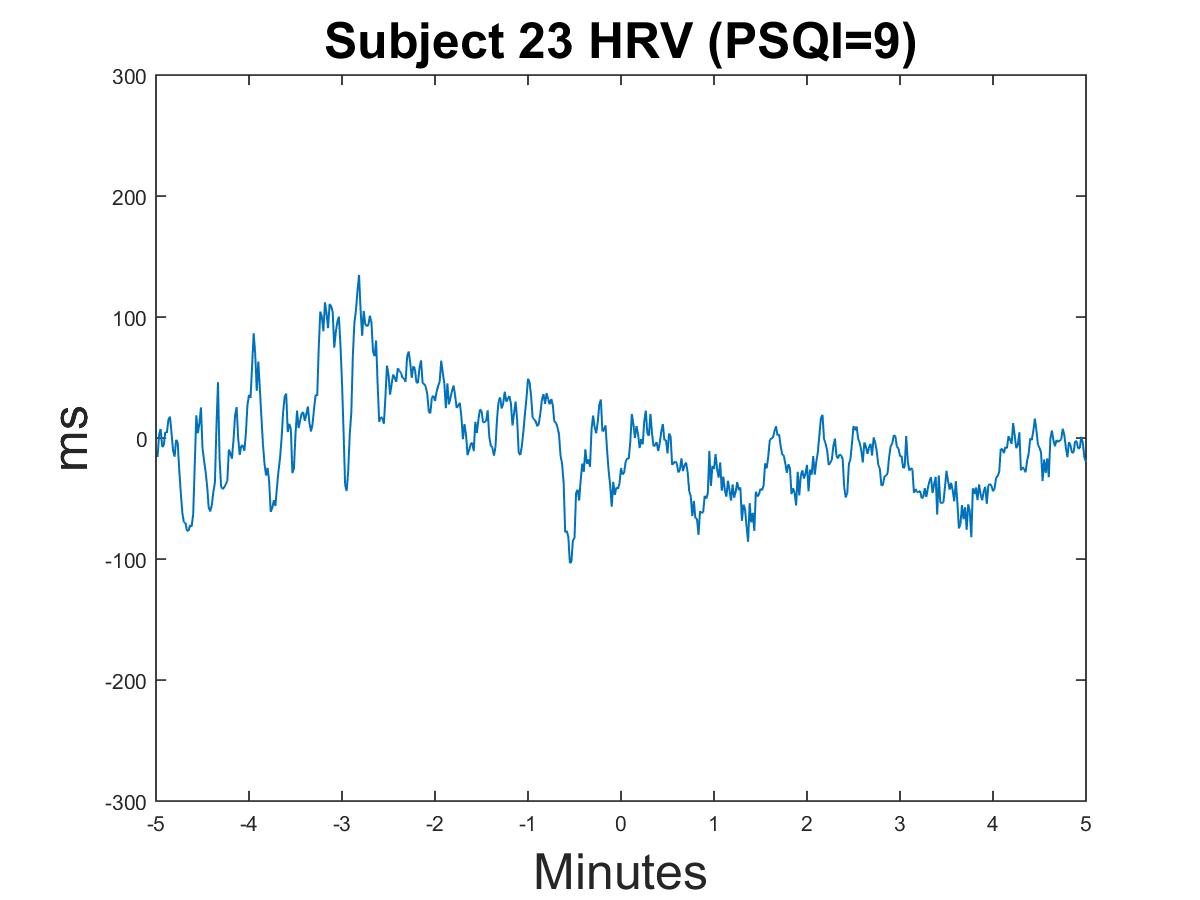}\includegraphics[width=0.5\textwidth]{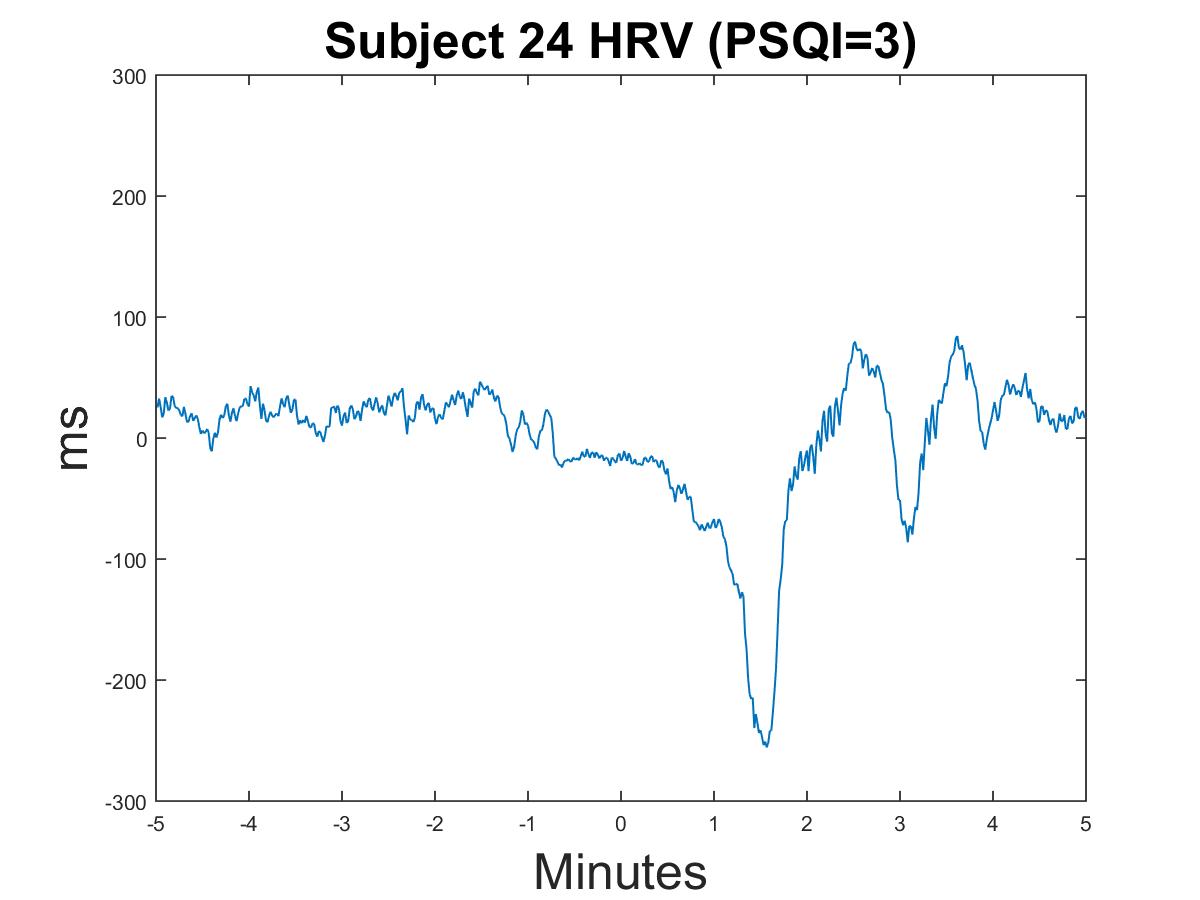}\\
\vspace{2cm}
\includegraphics[width=0.5\textwidth]{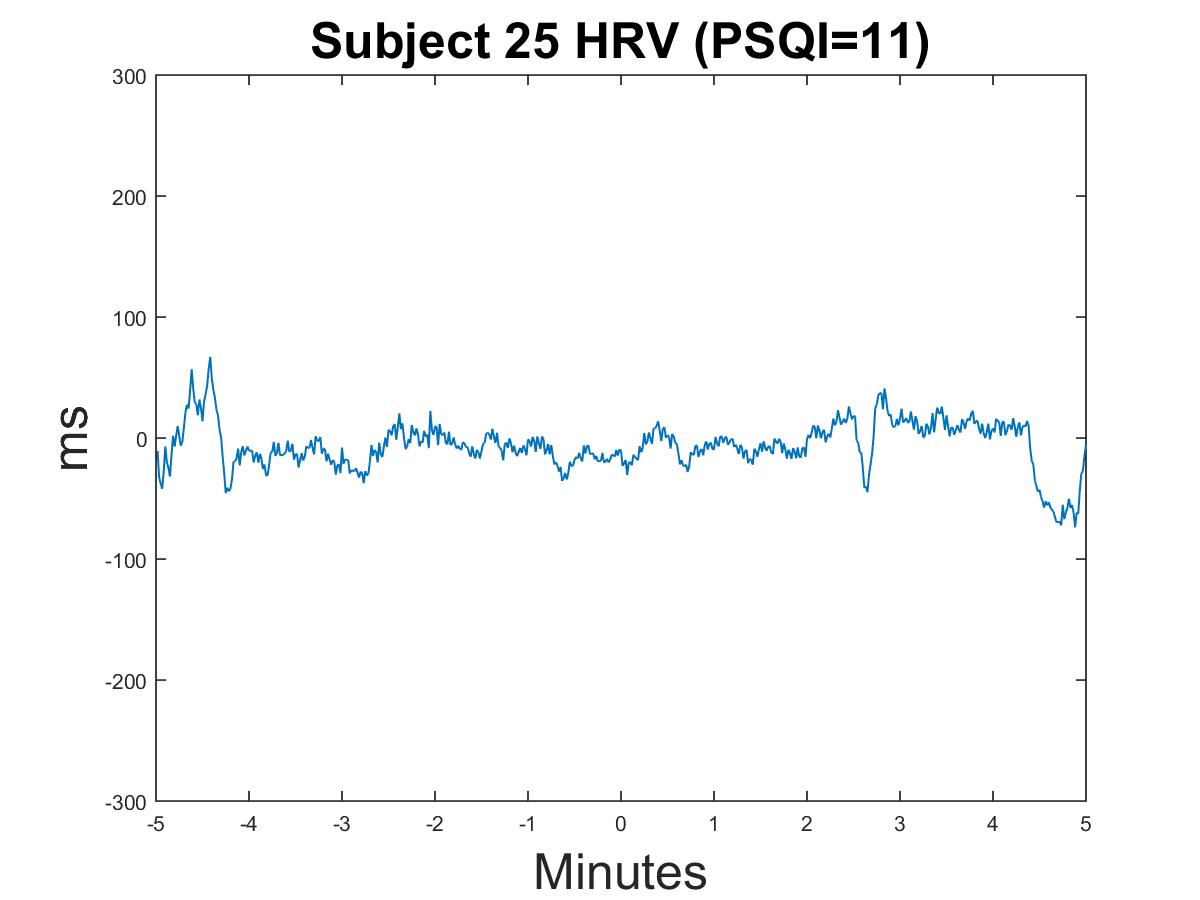}\includegraphics[width=0.5\textwidth]{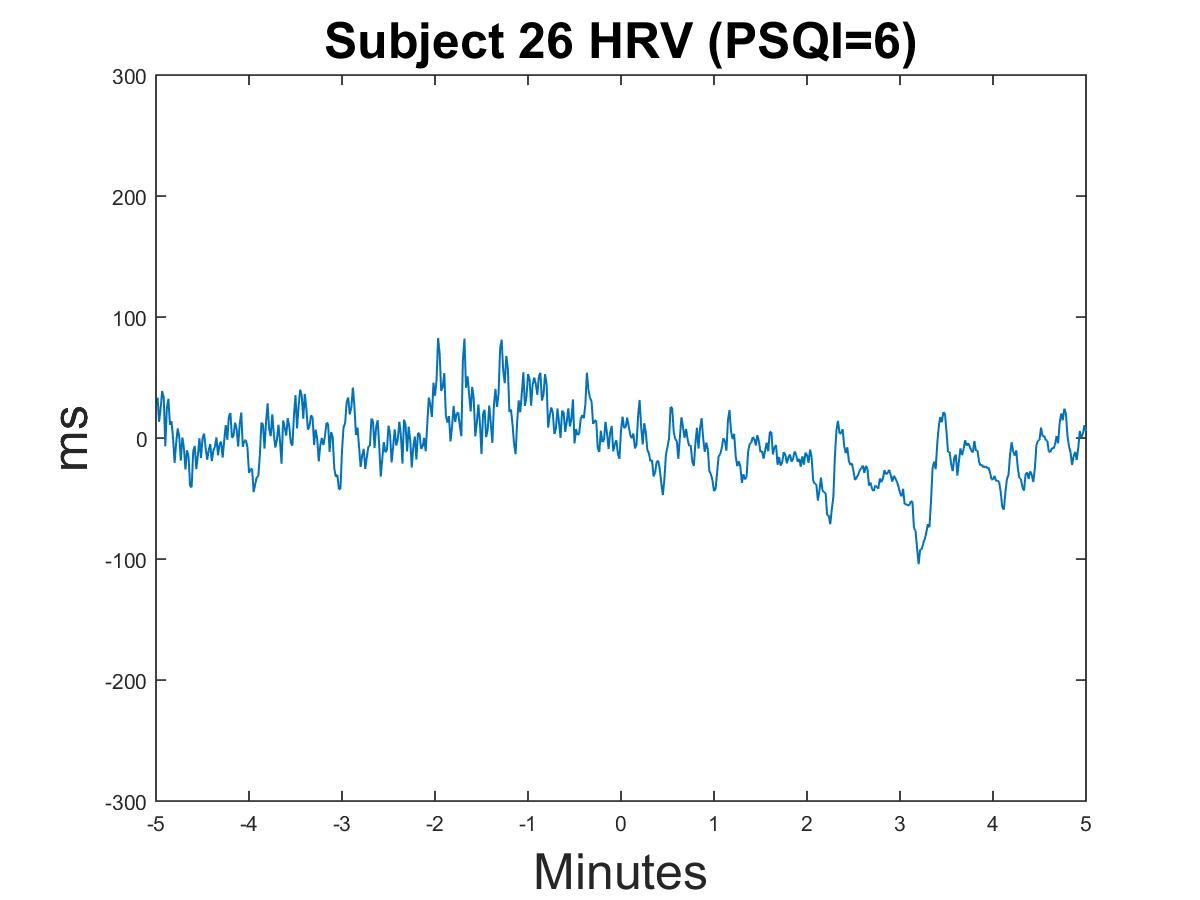}\\
\vspace{2cm}
\includegraphics[width=0.5\textwidth]{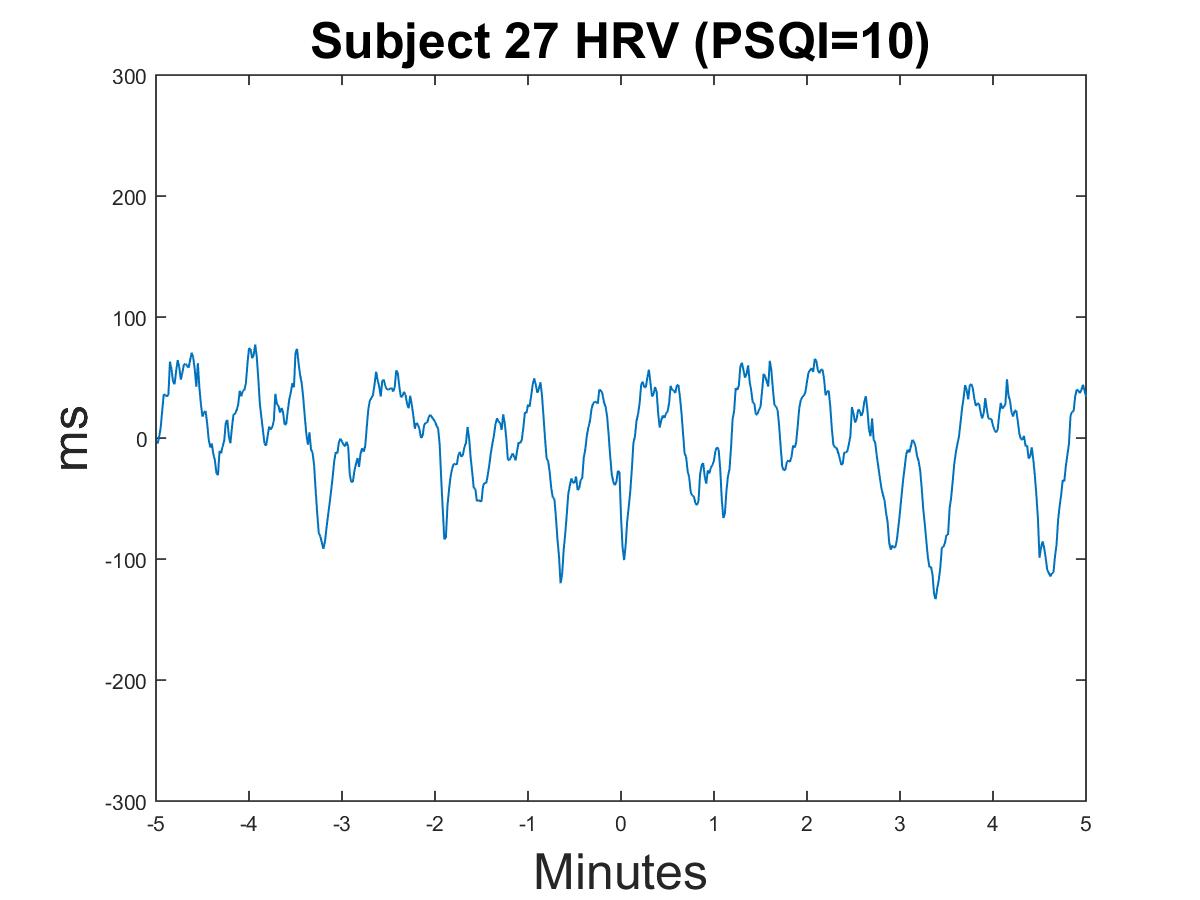}\includegraphics[width=0.5\textwidth]{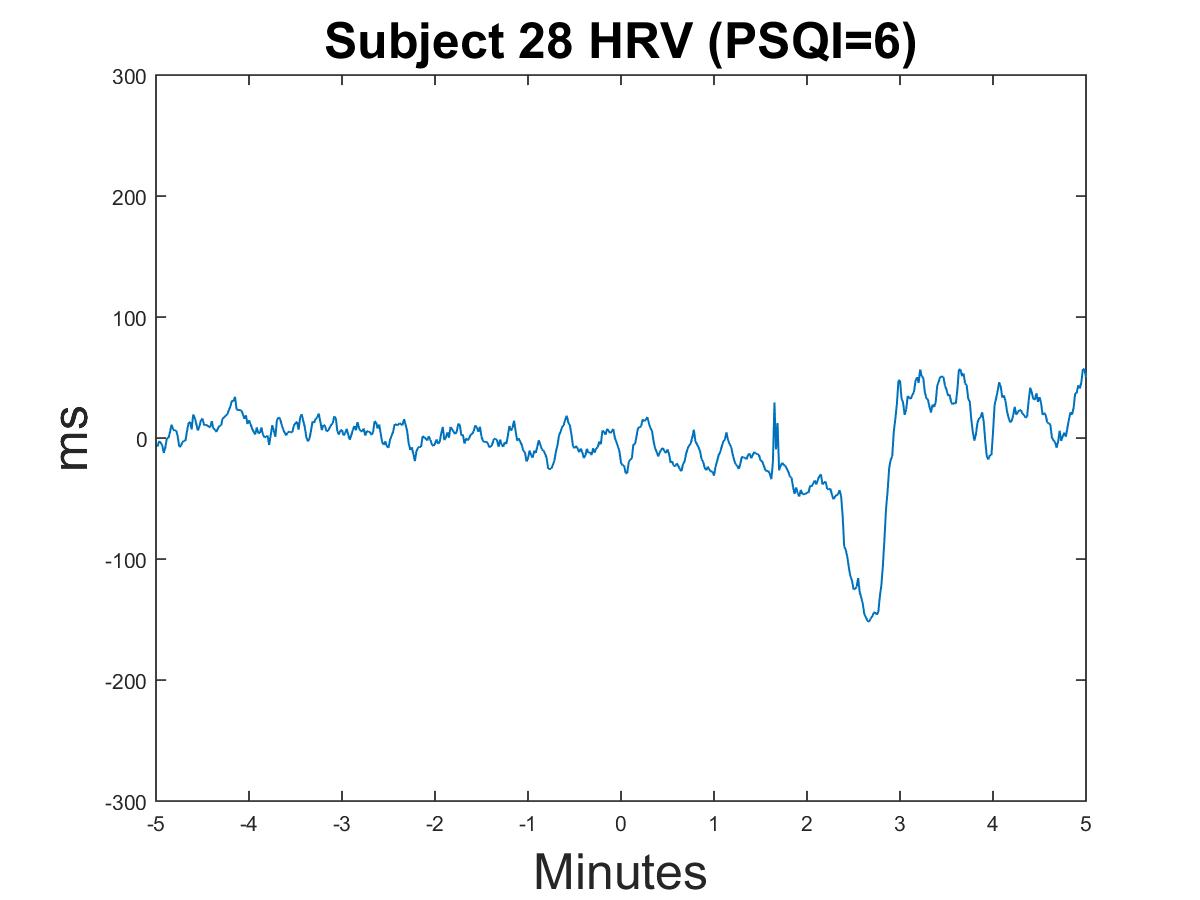}\\

\includegraphics[width=0.5\textwidth]{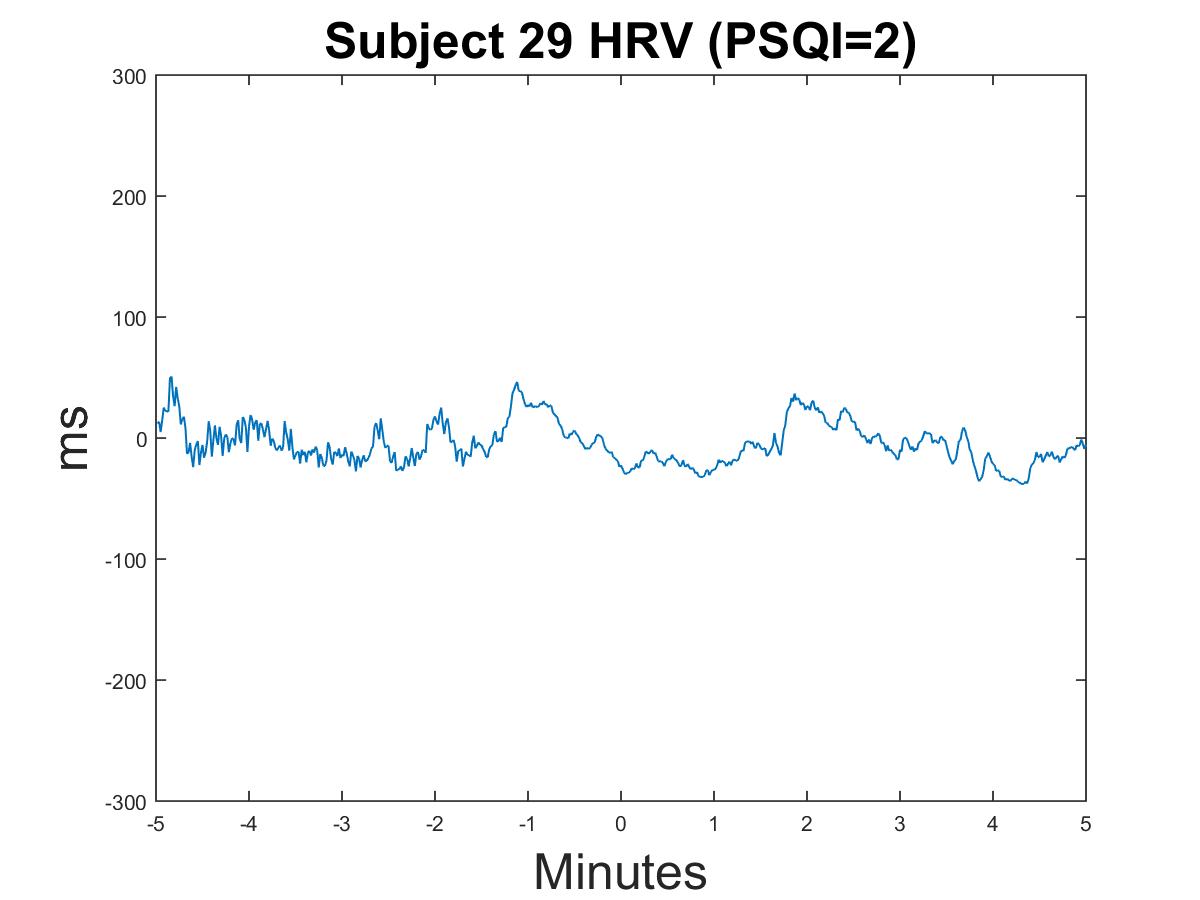}\includegraphics[width=0.5\textwidth]{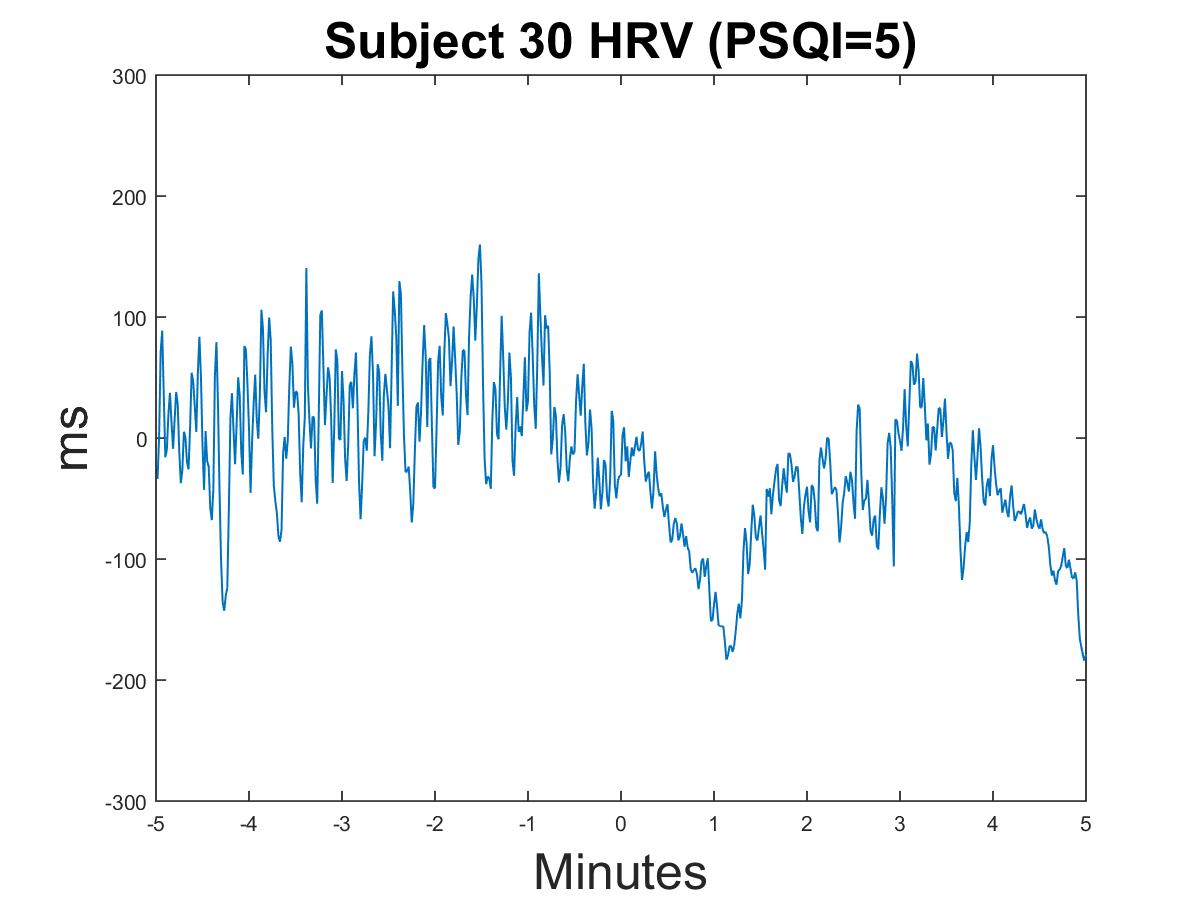}\\
\end{center}

\section*{Web Supplement B: Sampling Scheme Details}
\label{sec:websuppB}

Let the time-covariate plane of values be partitioned into $m$ time-based segments and $p$ covariate-based segments and denoted by $\boldsymbol{\xi}_m = (\xi_{0,m},\ldots,\xi_{m,m})$ and $\boldsymbol{\psi}_p = (\psi_{0,p},\ldots,\psi_{p,p})$ respectively.  Also let $\boldsymbol{\tau}^2_{m,p}$  be the $m \times p$ matrix of smoothing parameters and 
$\boldsymbol{\beta}_{m,p}$ be the $m \times p$ array of spline coefficients 
such that $\{\tau^2\}_{i,j}$ is the smoothing parameter and $\{\boldsymbol{\beta}\}_{i,j}$ is a vector of unknown coefficients for the block defined by the $i$th time-based segment and $j$th covariate-based segment for $i=1,\ldots,m$ and $j=1,\ldots,p$.  As previously mentioned, every iteration of the MCMC estimation procedure evaluates the following moves in order:
\begin{enumerate}
\item Time-based between-model move
\item Time-based within-model move
\item Covariate-based between-model move
\item Covariate-based within-model move
\end{enumerate}
where the details of each move are contained in the following sections.

\subsection*{1. Time-Based Between-Model Moves}
Let $\boldsymbol{\theta}_{m,p} = (\boldsymbol{\tau}^2_{m,p}, \boldsymbol{\beta}_{m,p})$, and suppose the MCMC estimation is currently at $(m^c,p^c, \boldsymbol{\xi}^c_{m^c}, \boldsymbol{\psi}^c_{p^c}, \boldsymbol{\theta}^c_{m^c,p^c})$.  Since we are considering a move in the time-based segmentation, all of the densities are conditional on $(p^c, \boldsymbol{\psi}^c_{p^c})$ which have been suppressed here for ease of notation.  We propose a move to $(m^p,p^c, \boldsymbol{\xi}^p_{m^p}, \boldsymbol{\psi}^c_{p^c}, \boldsymbol{\theta}^p_{m^p,p^c})$ by drawing from a proposal density $q(m^p,\boldsymbol{\xi}^p_{m^p},\boldsymbol{\theta}^p_{m^p,p^c}|m^c,\boldsymbol{\xi}^c_{m^c},\boldsymbol{\theta}^c_{m^c,p^c})$ and accepting with probability

$$\alpha_m = \mathrm{min}\left\{1,\frac{p(m^p,\boldsymbol{\xi}^p_{m^p},\boldsymbol{\theta}^p_{m^p,p^c}|\mathbf{x}) \times q(m^c,\boldsymbol{\xi}^c_{m^c},\boldsymbol{\theta}^c_{m^c,p^c}|m^p,\boldsymbol{\xi}^p_{m^p},\boldsymbol{\theta}^p_{m^p,p^c})}{p(m^c,\boldsymbol{\xi}^c_{m^c},\boldsymbol{\theta}^c_{m^c,p^c}|\mathbf{x}) \times q(m^p,\boldsymbol{\xi}^p_{m^p},\boldsymbol{\theta}^p_{m^p,p^c}|m^c,\boldsymbol{\xi}^c_{m^c},\boldsymbol{\theta}^c_{m^c,p^c})}\right\}.$$

The proposal density can be constructed as

\begin{align*}
q&(m^p,\boldsymbol{\xi}^p_{m^p},\boldsymbol{\theta}^p_{m^p,p^c}|m^c,\boldsymbol{\xi}^c_{m^c},\boldsymbol{\theta}^c_{m^c,p^c}) \\ 
&= q(m^p|m^c) \times q(\boldsymbol{\xi}^p_{m^p}| m^p, m^c, \boldsymbol{\xi}^c_{m^c}, \boldsymbol{\theta}^c_{m^c,p^c}) \\ 
&\times q(\boldsymbol{\tau}^{2p}_{m^p,p^c}| \boldsymbol{\xi}^p_{m^p}, m^p, m^c, \boldsymbol{\xi}^c_{m^c}, \boldsymbol{\theta}^c_{m^c,p^c}) \\ 
&\times q(\boldsymbol{\beta}^{p}_{m^p,p^c}| \boldsymbol{\tau}^{2p}_{m^p,p^c}, \boldsymbol{\xi}^p_{m^p}, m^p, m^c,\boldsymbol{\xi}^c_{m^c}, \boldsymbol{\theta}^c_{m^c,p^c}).
\end{align*}

So we have that $m^p$ is drawn first, followed by $\boldsymbol{\xi}^p_{m^p}$, $\boldsymbol{\tau}^{2p}_{m^p,p^c}$, and then $\boldsymbol{\beta}^{p}_{m^p,p^c}$.  

\begin{enumerate}[label=(\alph*)]
\item Let $M$ be the maximum number of time-based segments allowed and $m_{2\mathrm{min}}^c$ be the current number of segments that contain at least $2t_{\mathrm{min}}$ observations. Then we have
$$q(m^p=k|m^c) = 
\begin{cases}
1, \qquad \mathrm{if} \; k=m^c-1 \; \mathrm{and} \; m^c=M \; \mathrm{or} \; m_{2\mathrm{min}}^c=0, \\
1, \qquad \mathrm{if} \; k=m^c+1 \; \mathrm{and} \; m^c=1, \\
1/2, \quad \mathrm{otherwise.}
\end{cases}$$

\item Given $m^p$, a new time-based partition, smoothing parameters, and coefficients are proposed as follows.
\begin{enumerate}[label=\roman*.]
\item \textbf{Birth} ($m^p=m^c+1$)

\begin{enumerate}[label=\Alph*.]
\item A partition,

$$\boldsymbol{\xi}^p_{m^p} = (\xi_{0,m^c}^c,\ldots,\xi_{k^*-1,m^c}^c,\xi_{k^*,m^p}^p,\xi_{k^*,m^c}^c,\ldots,\xi_{m^c,m^c}^c),$$

is proposed by first randomly selecting a segment $j=k^*$ for splitting among segments containing at least $2t_{\mathrm{min}}$ observations.  Then a point $t^*$ is randomly selected subject to the constraint $\xi_{k^*-1,m^c}^c + t_{\mathrm{min}} \le t^* \le \xi_{k^*-1,m^c}^c - t_{\mathrm{min}}.$  The resulting proposal density is

$$q(\boldsymbol{\xi}^p_{m^p}| m^p, m^c, \boldsymbol{\xi}^c_{m^c}) = \frac{1}{m^c_{2\mathrm{min}}} \times \frac{1}{\xi_{k^*,m^c}^c-\xi_{k^*-1,m^c}^c-2t_{\mathrm{min}}+1}$$

\item \label{itm:smooth} New smoothing parameters across all covariate-based segments
$$\boldsymbol{\tau}^{2p}_{j} = (\tau^{2c}_{1,j},\ldots,\tau^{2c}_{k^*-1,j},\tau^{2p}_{k^*,j},\tau^{2p}_{k^*+1,j},\tau^{2c}_{k^*+1,j},\ldots,\tau^{2c}_{m^c,j})'$$
for $j=1,\ldots,p^c$ are proposed following \cite{adaptspec} as
$$\tau^{2p}_{k^*,j} = \tau^{2c}_{k^*,j} \times \frac{u_j}{1-u_j},$$
$$\tau^{2p}_{k^*+1,j} = \tau^{2c}_{k^*+1,j} \times \frac{1-u_j}{u_j},$$
where $u_j$ is drawn from a standard uniform distribution.

\item \label{itm:coeff1}
New vectors of coefficients across all covariate-based segments
$$\boldsymbol{\beta}^{p}_{j} = (\boldsymbol{\beta}^{c}_{1,j},\ldots,\boldsymbol{\beta}^{c}_{k^*-1,j},\boldsymbol{\beta}^{p}_{k^*,j},\boldsymbol{\beta}^{p}_{k^*+1,j},\boldsymbol{\beta}^{c}_{k^*+1,j},\ldots,\boldsymbol{\beta}^{c}_{m^c,j})'$$

for $j=1,\ldots,p^c$ are proposed by drawing pairs of vectors $\boldsymbol{\beta}^{p}_{k^*,j}$ and $\boldsymbol{\beta}^{p}_{k^*+1,j})$ from normal approximations to their posterior conditional distributions $p(\boldsymbol{\beta}^{p}_{k^*,j}|\mathbf{x}_{k^*,j}^p,\tau^{2p}_{k^*,j},m^p,p^c)$ and $p(\boldsymbol{\beta}^{p}_{k^*+1,j}|\mathbf{x}_{k^*+1,j}^p,\tau^{2p}_{k^*+1,j},m^p,p^c)$ where $\mathbf{x}_{k^*,j}^p$ and $\mathbf{x}_{k^*+1,j}^p$ represent the subsets of all replicated time series belonging to covariate-based segment $j$ and time-based segment $k^*$ and $k^*+1$ respectively (see Equation 1 in article).  As an example, $\boldsymbol{\beta}^{p}_{k^*,j}$ is drawn from $N(\boldsymbol{\beta}^{\mathrm{max}}_{k^*,j},\Sigma^{\mathrm{max}}_{k^*,j})$ where, 

$$\boldsymbol{\beta}^{\mathrm{max}}_{k^*,j} = \mathrm{arg} \; \max\limits_{\boldsymbol{\beta}^{p}_{k^*,j}} p(\boldsymbol{\beta}^{p}_{k^*,j}|\mathbf{x}_{k^*,j}^p,\tau^{2p}_{k^*,j},m^p,p^c)
\; \mathrm{and}$$

\begin{multline*}
\Sigma^{\mathrm{max}}_{k^*,j} = \left\{ -[\partial^2 \log p(\boldsymbol{\beta}^{p}_{k^*,j}|\mathbf{x}_{k^*,j}^p,\tau^{2p}_{k^*,j},m^p,p^c) ] \right. \\
\left. / [\partial \boldsymbol{\beta}^{p}_{k^*,j} \partial \boldsymbol{\beta}^{p'}_{k^*,j}]|_{\boldsymbol{\beta}^{p}_{k^*,j} = \boldsymbol{\beta}^{\mathrm{max}}_{k^*,j}}\right\}^{-1}.
\end{multline*}

The acceptance probability for the birth move is $\alpha_m = \min(1,A_m)$, where

\begin{align*}
A_m= &\frac{p(\mathbf{x}|\boldsymbol{\xi}^p_{m^p},\boldsymbol{\theta}_{m^p,p^c}^p,m^p)p(\boldsymbol{\xi}^p_{m^p},\boldsymbol{\theta}_{m^p,p^c}^p|m^p)p(m^p)}{p(\mathbf{x}|\boldsymbol{\xi}^c_{m^c},\boldsymbol{\theta}_{m^c,p^c}^c,m^c)p(\boldsymbol{\xi}^c_{m^c},\boldsymbol{\theta}_{m^c,p^c}^c|m^c)p(m^c)} \\
&\times \frac{p(m^c|m^p) \prod_j p(\boldsymbol{\beta}^{c}_{k^*,j})}{p(m^p|m^c)p(\xi^p_{k^*,m^p}|m^p,m^c)\prod_jp(u_j) p(\boldsymbol{\beta}^{p}_{k^*,j})p(\boldsymbol{\beta}^{p}_{k^*+1,j})}\\
&\times \prod_j \left|\frac{\partial(\tau^{2p}_{k^*,j},\tau^{2p}_{k^*+1,j})}{\partial(\tau^{2c}_{k^*,j},u_j)}\right|,
\end{align*}

where $p(u_j)=1$  and the Jacobian is

$$\left|\frac{\partial(\tau^{2p}_{k^*,j},\tau^{2p}_{k^*+1,j})}{\partial(\tau^{2c}_{k^*,j},u_j)}\right| = \frac{2\tau^{2c}_{k^*,j}}{u_j(1-u_j)}=2\left(\tau^{2p}_{k^*,j}+\tau^{2p}_{k^*+1,j}\right)^2,$$

for $j=1,\ldots,p^c$. If the move is accepted, then $m^c=m^p, \boldsymbol{\xi}^c_{m^c}=\boldsymbol{\xi}^p_{m^p},$ and $\boldsymbol{\theta}_{m^c,p^c}^c=\boldsymbol{\theta}_{m^p,p^c}^p$.

\end{enumerate}
\item \textbf{Death $(m^p=m^c-1)$}
\begin{enumerate}[label=\Alph*.]
\item A partition,

$$\boldsymbol{\xi}^p_{m^p} = (\xi_{0,m^c}^c,\ldots,\xi_{k^*-1,m^c}^c,\xi_{k^*+1,m^c}^c,\ldots,\xi_{m^c,m^c}^c),$$

is proposed by randomly selecting one of the $m^c-1$ partition points to drop with each one being equally likely, so that

$$q(\boldsymbol{\xi}^p_{m^p}| m^p, m^c, \boldsymbol{\xi}^c_{m^c}) = \frac{1}{m^c-1}.$$

\item  New smoothing parameters across all covariate-based segments
$$\boldsymbol{\tau}^{2p}_{j} = (\tau^{2c}_{1,j},\ldots,\tau^{2c}_{k^*-1,j},\tau^{2p}_{k^*,j},\tau^{2c}_{k^*+2,j},\ldots,\tau^{2c}_{m^c,j})'$$
for $j=1,\ldots,p^c$ are proposed by reversing corresponding smoothing parameter proposal step for the birth move [see \hyperref[itm:smooth]{Step 1 (b) i B}].

$$\tau^{2p}_{k^*,j} = \sqrt{\tau^{2c}_{k^*,j}\tau^{2c}_{k^*+1,j}}$$

\item New vectors of coefficients across all covariate-based segments
$$\boldsymbol{\beta}^{p}_{j} = (\boldsymbol{\beta}^{c}_{1,j},\ldots,\boldsymbol{\beta}^{c}_{k^*-1,j},\boldsymbol{\beta}^{p}_{k^*,j},\boldsymbol{\beta}^{c}_{k^*+2,j},\ldots,\boldsymbol{\beta}^{c}_{m^c,j})'$$

for $j=1,\ldots,p^c$ are proposed by drawing from a normal approximation to its posterior distribution as detailed in the corresponding coefficient proposal step for the birth move [see \hyperref[itm:coeff1]{Step 1 (b) i C}]. 
The acceptance probability is the inverse of the acceptance probability from the birth move.  If the move is accepted, then $m^c=m^p, \boldsymbol{\xi}^c_{m^c}=\boldsymbol{\xi}^p_{m^p},$ and $\boldsymbol{\theta}_{m^c,p^c}^c=\boldsymbol{\theta}_{m^p,p^c}^p$.

\end{enumerate}
\end{enumerate}
\end{enumerate}

\subsection*{2. Time-Based Within-Model Moves}
For this move, we first propose to relocate an existing time-based partition point and update the basis function coefficients across all covariate-based segments accordingly.  These two steps are jointly accepted or rejected in an M-H step.  Afterwards, smoothing parameters are updated in a Gibbs step.

\begin{enumerate}[label=(\alph*)]
\item Let $\boldsymbol{\theta} = (\boldsymbol{\tau}^2, \boldsymbol{\beta})$, and suppose the MCMC estimation is currently at $(\boldsymbol{\xi}^c, \boldsymbol{\psi}^c, \boldsymbol{\theta}^c)$.  We propose a move to $(\boldsymbol{\xi}^p, \boldsymbol{\psi}^c, \boldsymbol{\theta}^p)$ as follows.
\begin{enumerate}[label=\roman*.]
\item \label{itm:within1} Randomly select a partition point, $\xi_{k^*}$ to relocate from $m-1$ possible partition points so that

$$p(j=k^*) = 1/(m-1).$$

A new position on the interval $[\xi_{k^*-1},\xi_{k^*+1}]$ is then selected from a mixture distribution, following \cite{adaptspec}, so that

$$p(\xi_{k^*}^p=t|j=k^*) = \pi q_1(\xi_{k^*}^p = t|\xi_{k^*}^c) + (1-\pi) q_2(\xi_{k^*}^p = t|\xi_{k^*}^c),$$

where $q_1(\xi_{k^*}^p = t|\xi_{k^*}^c) = 1/(\xi_{k^*+1}^c - \xi_{k^*-1}^c - 2t_{\mathrm{min}}+1)$ for $\xi_{k^*-1}^c+t_{\mathrm{min}} \le t \le \xi_{k^*+1}^c-t_{\mathrm{min}}$, and
\begin{align*}
q_2(\xi_{k^*}^p &= t|\xi_{k^*}^c) = \\
&\begin{cases}
0, \; \; \; \, \quad \mathrm{if} \; |t-\xi_{k^*}^c| > 1,\\
1/3, \quad \mathrm{if} \; |t-\xi_{k^*}^c| \le 1 \; , \; \xi_{k^*+1}^c - \xi_{k^*}^c \ne t_{\mathrm{min}}, \\ 
\qquad \qquad \; \mathrm{and} \; \xi_{k^*}^c - \xi_{k^*-1}^c \ne t_{\mathrm{min}}\\
1/2, \quad \mathrm{if} \; t-\xi_{k^*}^c \le 1 \; , \; \xi_{k^*+1}^c - \xi_{k^*}^c = t_{\mathrm{min}}, \\ 
\qquad \qquad\; \mathrm{and} \; \xi_{k^*}^c - \xi_{k^*-1}^c \ne t_{\mathrm{min}}\\
1/2, \quad \mathrm{if} \; \xi_{k^*}^c-t \le 1 \; , \; \xi_{k^*+1}^c - \xi_{k^*}^c \ne t_{\mathrm{min}}, \\ 
\qquad \qquad \; \mathrm{and} \; \xi_{k^*}^c - \xi_{k^*-1}^c = t_{\mathrm{min}}\\
1, \; \; \; \, \quad \mathrm{if} \; t=\xi_{k^*}^c \; , \; \xi_{k^*+1}^c - \xi_{k^*}^c = t_{\mathrm{min}}, \\ 
\qquad \qquad\; \mathrm{and} \; \xi_{k^*}^c - \xi_{k^*-1}^c = t_{\mathrm{min}}\\
\end{cases}
\end{align*}

Note that $q_1(\xi_{k^*}^p = t|\xi_{k^*}^c)$ allows for bigger jumps to explore the parameter space more efficiently, while $q_2(\xi_{k^*}^p = t|\xi_{k^*}^c)$ moves at most one time point from the previous partition point.  The parameter $\pi$ allows for users to adjust the estimation procedure to find an appropriate balance between acceptance rates and efficient exploration of the parameter space.

\item New vectors of basis function coefficients across all covariate-based segments are proposed, $\boldsymbol{\beta}_{i,j}$ for $i=k^*,k^*+1$ and $j=1,\ldots,p^c$, from an approximation to $\prod_{i,j} p(\boldsymbol{\beta}_{i,j}|\mathbf{x}_{i,j}^p,\tau^2_{i,j}),$ as in the corresponding coefficient proposal step for the birth move [see \hyperref[itm:coeff1]{Step 1 (b) i C}].
The proposal density is then evaluated at the proposed and current values for the coefficients,

$$\prod_{i,j} q(\boldsymbol{\beta}^p_{i,j}|\mathbf{x}_{i,j}^p,\tau^2_{i,j}) \qquad \mathrm{and} \qquad \prod_{i,j} q(\boldsymbol{\beta}^c_{i,j}|\mathbf{x}_{i,j}^c,\tau^2_{i,j}).$$

and the move is accepted with probability

$$\alpha_m = \min \left\{1, \frac{\prod_{i,j}p(\mathbf{x}_{i,j}^p|\boldsymbol{\beta}^p_{i,j})p(\boldsymbol{\beta}^p_{i,j}|\tau^2_{i,j})
q(\boldsymbol{\beta}^c_{i,j}|\mathbf{x}_{i,j}^c,\tau^2_{i,j})}{\prod_{i,j}p(\mathbf{x}_{i,j}^c|\boldsymbol{\beta}^c_{i,j})p(\boldsymbol{\beta}^c_{i,j}|\tau^2_{i,j})
q(\boldsymbol{\beta}^p_{i,j}|\mathbf{x}_{i,j}^p,\tau^2_{i,j})} \right\}.$$

If the move is accepted, then $\xi_{k^*}^c = \xi_{k^*}^p$, and $(\boldsymbol{\beta}^c_{k^*,j}, \boldsymbol{\beta}^c_{k^*+1,j}) = (\boldsymbol{\beta}^p_{k^*,j}, \boldsymbol{\beta}^p_{k^*+1,j})$ for $j=1,\ldots,p^c$.

\end{enumerate}

\item Draw new smoothing parameters across all covariate-based segments from 

$$\prod_{i,j} p(\boldsymbol{\tau}^{2}_{i,j}|\boldsymbol{\beta}_{i,j})$$

for $i=k^*,k^*+1$ and $j=1,\ldots,p^c$.  Where the density for $p(\tau^2|\boldsymbol{\beta})$ is proportional to Equation 2 from article.
\end{enumerate}

\subsection*{3. Covariate-Based Between-Model Moves}
The target and proposal densities for moves made in the covariate-based segments are conditional on $(m^c,\boldsymbol{\xi}^c_{m^c})$, so that notation has been dropped.  The structure of the moves are similar to those made in the time-based segments, so much of the additional explanations provided in Steps 1 and 2 will not be reproduced.  Again, let $\boldsymbol{\theta}_{m,p} = (\boldsymbol{\tau}^2_{m,p}, \boldsymbol{\beta}_{m,p})$, and suppose the MCMC estimation is currently at $(p^c, \boldsymbol{\psi}^c_{p^c}, \boldsymbol{\theta}^c_{m^c,p^c})$.  We propose a move to $(p^p, \boldsymbol{\psi}^p_{p^p}, \boldsymbol{\theta}^p_{m^c,p^p})$ by drawing from a proposal density $q(p^p,\boldsymbol{\psi}^p_{p^p},\boldsymbol{\theta}^p_{m^c,p^p}|p^c,\boldsymbol{\psi}^c_{p^c},\boldsymbol{\theta}^c_{m^c,p^c})$ and accepting with probability

$$\alpha_p = \mathrm{min}\left\{1,\frac{p(p^p,\boldsymbol{\psi}^p_{p^p},\boldsymbol{\theta}^p_{m^c,p^p}|\mathbf{x}) \times q(p^c,\boldsymbol{\psi}^c_{p^c},\boldsymbol{\theta}^c_{m^c,p^c}|p^p,\boldsymbol{\psi}^p_{p^p},\boldsymbol{\theta}^p_{m^c,p^p})}{p(p^c,\boldsymbol{\psi}^c_{p^c},\boldsymbol{\theta}^c_{m^c,p^c}|\mathbf{x}) \times q(p^p,\boldsymbol{\psi}^p_{p^p},\boldsymbol{\theta}^p_{m^c,p^p}|p^c,\boldsymbol{\psi}^c_{p^c},\boldsymbol{\theta}^c_{m^c,p^c})}\right\}.$$

For the proposal density, we draw $p^p$ first, followed by $\boldsymbol{\psi}^p_{p^p}$, $\boldsymbol{\tau}^{2p}_{m^c,p^p}$, and then $\boldsymbol{\beta}^{p}_{m^c,p^p}$ as follows.  

\begin{enumerate}[label=(\alph*)]
\item Let $M_p$ be the maximum number of covariate-based segments allowed and $p_{2\mathrm{min}}^c$ be the current number of segments that contain at least $2w_{\mathrm{min}}$ realizations. Then we have
$$q(p^p=k|p^c) = 
\begin{cases}
1, \qquad \mathrm{if} \; k=p^c-1 \; \mathrm{and} \; p^c=M_p \; \mathrm{or} \; p_{2\mathrm{min}}^c=0, \\
1, \qquad \mathrm{if} \; k=p^c+1 \; \mathrm{and} \; p^c=1, \\
1/2, \quad \mathrm{otherwise.}
\end{cases}$$

\item Given $p^p$, smoothing parameters and coefficients are proposed as follows.
\begin{enumerate}[label=\roman*.]
\item \textbf{Birth} ($p^p=p^c+1$)

\begin{enumerate}[label=\Alph*.]
\item A partition,

$$\boldsymbol{\psi}^p_{p^p} = (\psi_{0,p^c}^c,\ldots,\psi_{k^*-1,p^c}^c,\psi_{k^*,p^p}^p,\psi_{k^*,p^c}^c,\ldots,\psi_{p^c,p^c}^c),$$

is proposed by randomly selecting a segment $j=k^*$ for splitting among segments containing at least $2w_{\mathrm{min}}$ realizations.  Then a point $w^*$ is randomly selected subject to the constraint that the resulting covariate-based segments contain at least $w_{\mathrm{min}}$ realizations in each segment.  Let $r_{k^*}$ be the number of distinct covariate values in segment $k^*$.  The resulting proposal density is

$$q(\boldsymbol{\psi}^p_{p^p}| p^p, p^c, \boldsymbol{\psi}^c_{p^c}) = \frac{1}{p^c_{2\mathrm{min}}} \times \frac{1}{r_{k^*}-2w_{\mathrm{min}}+1}$$

\item New smoothing parameters across all time-based segments
$$\boldsymbol{\tau}^{2p}_{j} = (\tau^{2c}_{j,1},\ldots,\tau^{2c}_{j,k^*-1},\tau^{2p}_{j,k^*},\tau^{2p}_{j, k^*+1},\tau^{2c}_{j,k^*+1},\ldots,\tau^{2c}_{j,p^c})$$
for $j=1,\ldots,m^c$ are proposed as
$$\tau^{2p}_{j,k^*} = \tau^{2c}_{j,k^*} \times \frac{u_j}{1-u_j},$$
$$\tau^{2p}_{j,k^*+1} = \tau^{2c}_{j,k^*+1} \times \frac{1-u_j}{u_j},$$
where $u_j$ is drawn from a standard uniform distribution.

\item
New vectors of coefficients across all time-based segments
$$\boldsymbol{\beta}^{p}_{j} = (\boldsymbol{\beta}^{c}_{j,1},\ldots,\boldsymbol{\beta}^{c}_{j,k^*-1},\boldsymbol{\beta}^{p}_{j,k^*},\boldsymbol{\beta}^{p}_{j,k^*+1},\boldsymbol{\beta}^{c}_{j,k^*+1},\ldots,\boldsymbol{\beta}^{c}_{j,p^c})$$

for $j=1,\ldots,m^c$ are proposed by drawing pairs of vectors $\boldsymbol{\beta}^{p}_{j,k^*}$ and $\boldsymbol{\beta}^{p}_{j,k^*+1})$ from normal approximations to their posterior conditional distributions as detailed in the corresponding time-based birth move coefficient proposal step [see \hyperref[itm:coeff1]{Step 1 (b) i C}].
The acceptance probability for the birth move is $\alpha_p = \min(1,A_p)$, where

\begin{align*}
A_p= &\frac{p(\mathbf{x}|\boldsymbol{\psi}^p_{p^p},\boldsymbol{\theta}_{m^c,p^p}^p,p^p)p(\boldsymbol{\psi}^p_{p^p},\boldsymbol{\theta}_{m^c,p^p}^p|p^p)p(p^p)}{p(\mathbf{x}|\boldsymbol{\psi}^c_{p^c},\boldsymbol{\theta}_{m^c,p^c}^c,p^c)p(\boldsymbol{\psi}^c_{p^c},\boldsymbol{\theta}_{m^c,p^c}^c|p^c)p(p^c)} \\
&\times \frac{p(p^c|p^p) \prod_j p(\boldsymbol{\beta}^{c}_{j,k^*})}{p(p^p|p^c)p(\psi^p_{k^*,p^p}|p^p,p^c)\prod_jp(u_j) p(\boldsymbol{\beta}^{p}_{j,k^*})p(\boldsymbol{\beta}^{p}_{j,k^*+1})}\\
&\times \prod_j \left|\frac{\partial(\tau^{2p}_{j,k^*},\tau^{2p}_{j,k^*+1})}{\partial(\tau^{2c}_{j,k^*},u_j)}\right|,
\end{align*}

where $p(u_j)=1$  and the Jacobian is

$$\left|\frac{\partial(\tau^{2p}_{j,k^*},\tau^{2p}_{j,k^*+1})}{\partial(\tau^{2c}_{j,k^*},u_j)}\right| = \frac{2\tau^{2c}_{j,k^*}}{u_j(1-u_j)}=2\left(\tau^{2p}_{j,k^*}+\tau^{2p}_{j,k^*+1}\right)^2,$$

for $j=1,\ldots,m^c$. If the move is accepted, then $p^c=p^p, \boldsymbol{\psi}^c_{p^c}=\boldsymbol{\psi}^p_{p^p},$ and $\boldsymbol{\theta}_{m^c,p^c}^c=\boldsymbol{\theta}_{m^c,p^p}^p$.

\end{enumerate}
\item \textbf{Death $(p^p=p^c-1)$}
\begin{enumerate}[label=\Alph*.]
\item A partition,

$$\boldsymbol{\psi}^p_{p^p} = (\psi_{0,p^c}^c,\ldots,\psi_{k^*-1,p^c}^c,\psi_{k^*+1,p^c}^c,\ldots,\psi_{p^c,p^c}^c),$$

is proposed by randomly selecting one of the $p^c-1$ partition points to drop, so that

$$q(\boldsymbol{\psi}^p_{p^p}| p^p, p^c, \boldsymbol{\psi}^c_{p^c}) = \frac{1}{p^c-1}.$$

\item  New smoothing parameters across all time-based segments
$$\boldsymbol{\tau}^{2p}_{j} = (\tau^{2c}_{j,1},\ldots,\tau^{2c}_{j,k^*-1},\tau^{2p}_{j,k^*},\tau^{2c}_{j,k^*+2},\ldots,\tau^{2c}_{j,p^c})$$
for $j=1,\ldots,m^c$ are proposed similar to step 1 (b) ii B:

$$\tau^{2p}_{j,k^*} = \sqrt{\tau^{2c}_{j,k^*}\tau^{2c}_{j,k^*+1}}$$

\item New vectors of coefficients across all time-based segments
$$\boldsymbol{\beta}^{p}_{j} = (\boldsymbol{\beta}^{c}_{j,1},\ldots,\boldsymbol{\beta}^{c}_{j,k^*-1},\boldsymbol{\beta}^{p}_{j,k^*},\boldsymbol{\beta}^{c}_{j,k^*+2},\ldots,\boldsymbol{\beta}^{c}_{j,m^c})$$

for $j=1,\ldots,m^c$ are proposed by drawing from a normal approximation to its posterior distribution as detailed in the corresponding time-based birth move coefficient proposal step [see \hyperref[itm:coeff1]{Step 1 (b) i C}].
The acceptance probability is the inverse of the acceptance probability from the birth move.  If the move is accepted, then $p^c=p^p, \boldsymbol{\psi}^c_{p^c}=\boldsymbol{\psi}^p_{p^p},$ and $\boldsymbol{\theta}_{m^c,p^c}^c=\boldsymbol{\theta}_{m^c,p^p}^p$.

\end{enumerate}
\end{enumerate}
\end{enumerate}

\subsection*{4. Covariate-Based Within-Model Moves}

For this move, we first propose to relocate an existing covariate-based partition point and update the basis function coefficients across all time-based segments accordingly.  These two steps are jointly accepted or rejected in an M-H step.  Afterwards, smoothing parameters are updated in a Gibbs step.

\begin{enumerate}[label=(\alph*)]
\item Let $\boldsymbol{\theta} = (\boldsymbol{\tau}^2, \boldsymbol{\beta})$, and suppose the MCMC estimation is currently at $(\boldsymbol{\xi}^c, \boldsymbol{\psi}^c, \boldsymbol{\theta}^c)$.  We propose a move to $(\boldsymbol{\xi}^c, \boldsymbol{\psi}^p, \boldsymbol{\theta}^p)$ as follows.
\begin{enumerate}[label=\roman*.]
\item Randomly select a partition point, $\psi_{k^*}$ to relocate from $p-1$ possible partition points so that

$$p(j=k^*) = 1/(p-1).$$

A new position on the interval $[\psi_{k^*-1},\psi_{k^*+1}]$ is then selected from a mixture distribution similar to the corresponding time-based within-model move step [see \hyperref[itm:within1]{Step 2 (a) i}],
so that

$$p(\psi_{k^*}^p=w|j=k^*) = \pi_p q_1(\psi_{k^*}^p = w|\psi_{k^*}^c) + (1-\pi_p) q_2(\psi_{k^*}^p = w|\psi_{k^*}^c),$$

where $q_1(\psi_{k^*}^p = w|\psi_{k^*}^c) = 1/(r_{k^*}+r_{k^*+1} - 2w_{\mathrm{min}}+1)$, and

\begin{align*}
q_2(\psi_{k^*}^p &= w|\psi_{k^*}^c) = \\
&\begin{cases}
0, \; \; \; \, \quad \mathrm{if} \; w<\psi_{k^*}^{c-} \; \mathrm{or} \; w>\psi_{k^*}^{c+},\\
1/3, \quad \mathrm{if} \; w \in (\psi_{k^*}^{c-},\psi_{k^*}^{c},\psi_{k^*}^{c+}) \; , \; r_{k^*+1} \ne w_{\mathrm{min}}, \\ 
\qquad \qquad \; \mathrm{and} \; r_{k^*} \ne w_{\mathrm{min}}\\
1/2, \quad \mathrm{if} \; w \in (\psi_{k^*}^{c-},\psi_{k^*}^{c}) \; , \; r_{k^*+1} = w_{\mathrm{min}}, \\ 
\qquad \qquad\; \mathrm{and} \; r_{k^*} \ne w_{\mathrm{min}}\\
1/2, \quad \mathrm{if} \; w \in (\psi_{k^*}^{c},\psi_{k^*}^{c+}) \; , \; r_{k^*+1} \ne w_{\mathrm{min}}, \\ 
\qquad \qquad \; \mathrm{and} \; r_{k^*} = w_{\mathrm{min}}\\
1, \; \; \; \, \quad \mathrm{if} \; w=\psi_{k^*}^{c} \; , \; r_{k^*+1} = w_{\mathrm{min}}, \\ 
\qquad \qquad\; \mathrm{and} \; r_{k^*} = w_{\mathrm{min}}\\
\end{cases}
\end{align*}
where $\psi_{k^*}^{c-}$ is the largest distinct covariate value less than $\psi_{k^*}^c$, $\psi_{k^*}^{c+}$ is the smallest distinct covariate value greater than $\psi_{k^*}^c$, and $r_{k^*}$ and $r_{k^*+1}$ are the number of realizations in covariate-based segment $k^*$ and $k^*+1$ respectively.

\item New vectors of basis function coefficients across all time-based segments are proposed, $\boldsymbol{\beta}_{i,j}$ for $i=1,\ldots,m^c$ and $j=k^*,k^*+1$, from an approximation to $\prod_{i,j} p(\boldsymbol{\beta}_{i,j}|\mathbf{x}_{i,j}^p,\tau^2_{i,j}),$ as detailed in the corresponding time-based birth move coefficient proposal step [see \hyperref[itm:coeff1]{Step 1 (b) i C}]. 
The proposal density is then evaluated at the proposed and current values for the coefficients, and the move is accepted with probability

$$\alpha_p = \min \left\{1, \frac{\prod_{i,j}p(\mathbf{x}_{i,j}^p|\boldsymbol{\beta}^p_{i,j})p(\boldsymbol{\beta}^p_{i,j}|\tau^2_{i,j})
q(\boldsymbol{\beta}^c_{i,j}|\mathbf{x}_{i,j}^c,\tau^2_{i,j})}{\prod_{i,j}p(\mathbf{x}_{i,j}^c|\boldsymbol{\beta}^c_{i,j})p(\boldsymbol{\beta}^c_{i,j}|\tau^2_{i,j})
q(\boldsymbol{\beta}^p_{i,j}|\mathbf{x}_{i,j}^p,\tau^2_{i,j})} \right\}.$$

If the move is accepted, then $\psi_{k^*}^c = \psi_{k^*}^p$, and $(\boldsymbol{\beta}^c_{j,k^*}, \boldsymbol{\beta}^c_{j,k^*+1}) = (\boldsymbol{\beta}^p_{j,k^*}, \boldsymbol{\beta}^p_{j,k^*+1})$ for $j=1,\ldots,m^c$.

\end{enumerate}

\item Draw new smoothing parameters across all time-based segments from 

$$\prod_{i,j} p(\boldsymbol{\tau}^{2}_{i,j}|\boldsymbol{\beta}_{i,j})$$

for $i=1,\ldots,m^c$ and $j=k^*,k^*+1$.  Where the density for $p(\tau^2|\boldsymbol{\beta})$ is proportional to Equation 2 from article.
\end{enumerate}

\label{lastpage}

\end{document}